\documentclass[journal]{IEEEtran}

\usepackage{amsmath} 
\usepackage{amssymb}  
\usepackage{bm}  
\usepackage{dsfont}
\usepackage{graphicx,color} 
\usepackage{epsfig} 
\usepackage{color}
\usepackage[tight,footnotesize]{subfigure}
\usepackage{cite}
\usepackage{url}
\usepackage{array}
\usepackage[svgnames,table]{xcolor}
\usepackage{dpfloat, booktabs}
\usepackage{pdflscape}
\usepackage{rotating}
\usepackage{longtable}
\usepackage{booktabs}
\usepackage{multirow}
\usepackage{algpseudocode}
\usepackage{algorithm}
\usepackage[colorlinks,linkcolor=Black,citecolor=Black,urlcolor=Black,breaklinks=true]{hyperref}

\DeclareMathOperator*{\E}{{\mathbf E}}        
\let\Pr\undefined 
\DeclareMathOperator{\Pr}{{\mathbf P}}        

\newcommand{\vect}[1]{\bm{#1}} 

\newtheorem{theorem}{Theorem}[section]
\newtheorem{corollary}[theorem]{Corollary}
\newtheorem{lemma}[theorem]{Lemma}

\newtheorem{definition}{Definition}[section]

\newtheorem{assumption}{Assumption}[section]
\newtheorem{example}{Example}[section]

\newtheorem{remark}{Remark}[section]

\usepackage{epstopdf}
\graphicspath{{figuresTR/}}
\begin{document}

\title{Performance Analysis of \\
a Network of Event-based Systems}%

\author{Chithrupa~Ramesh,~\IEEEmembership{Student~Member,~IEEE,}
        Henrik~Sandberg,~\IEEEmembership{Member,~IEEE,}
        and~Karl~H.~Johansson,~\IEEEmembership{Fellow,~IEEE}
\thanks{This work was supported by the Swedish Research Council,
        VINNOVA (The Swedish Governmental Agency for Innovation
        Systems), the Swedish Foundation for Strategic Research,
        the Knut and Alice Wallenberg Foundation and the EU
        project Hycon$2$.}
\thanks{C. Ramesh, H. Sandberg and Karl H. Johansson are with the ACCESS Linnaeus Centre,
Electrical Engineering, KTH Royal Institute of Technology, Stockholm,
Sweden. e-mail: \{cramesh,hsan,kallej\}@kth.se.}
}
\maketitle


\begin{abstract}

We consider a scenario where multiple event-based systems use a wireless network to communicate with their respective controllers. These systems use a contention resolution mechanism (CRM) to arbitrate access to the network. We present a Markov model for the network interactions between the event-based systems. Using this model, we obtain an analytical expression for the reliability, or the probability of successfully transmitting a packet, in this network. There are two important aspects to our model. Firstly, our model captures the joint interactions of the event-triggering policy and the CRM. This is required because event-triggering policies typically adapt to the CRM outcome. Secondly, the model is obtained by decoupling interactions between the different systems in the network, drawing inspiration from Bianchi's analysis of IEEE~$802.11$. This is required because the network interactions introduce a correlation between the system variables. We present Monte-Carlo simulations that validate our model under various network configurations, and verify our performance analysis as well.

\end{abstract}

\section{Introduction} \label{S:Intro}

\subsection{Motivation}

\begin{figure*}[tb]
        \centering
        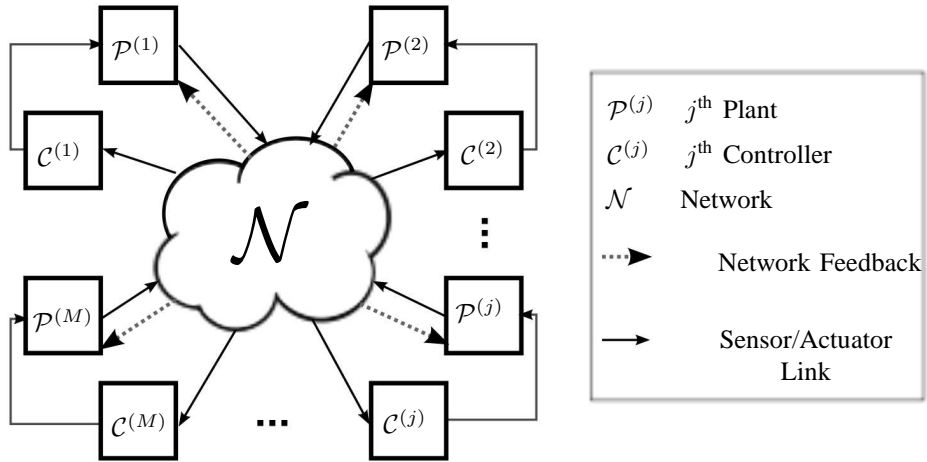
        \caption[NCSs with a shared sensing link]{A network of $M$ closed-loop systems, with each loop consisting of a plant $\mathcal{P}^{j}$ and a controller $\mathcal{C}^{j}$ for $j \in \{1,\dots,M\}$. The systems share access to a common medium on the sensor link, and adapt their traffic rates to the feedback from the network. The controllers and actuators communicate over dedicated links.} \label{Fig:macNCS}
\end{figure*}

Digital control systems often use the time-triggered paradigm, where a measurement is periodically sent to the controller to generate a control signal. Event-based systems provide an alternative, wherein only measurements that qualify as `events' are sent to the controller. These systems could result in fewer transmissions \cite{Astrom1999,Otanez2002}, which is an important consideration when multiple closed-loop systems use a shared network to communicate with their respective controllers. Many wireless networked control systems operate in this manner, as shown in Fig.~\ref{Fig:macNCS}, where wireless links connect sensors with controllers while the controller-actuator links are wired. The shared network may be able to support more number of event-based systems than time-triggered ones, with comparable system performances. To achieve this target, one must understand the interaction of multiple event-based systems in a shared network. Only then will it be possible to predict their performance, and design event-triggering policies that are matched to the available network resources. 

\begin{figure*}[tb]
        \centering
        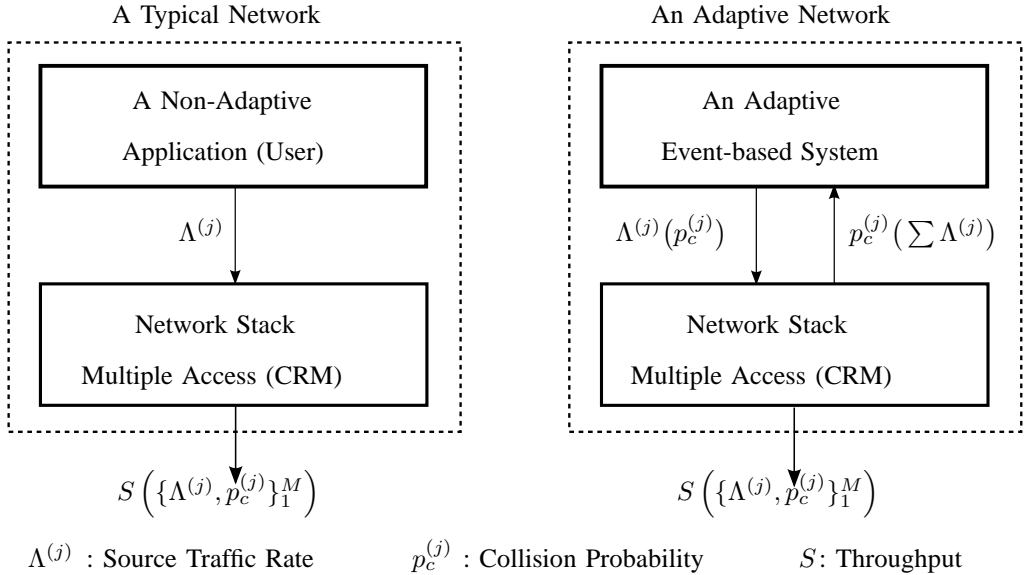
        \caption[Need for Joint Analysis in Event-based Networks]{A typical network can be analyzed by evaluating the traffic rate of the user, and the collision probability of the multiple access protocol, in isolation. In an event-based network, the input traffic is adapted to the traffic in the network, i.e., $\Lambda^{(j)}$ is a function of $p_{c}^{(j)}$. The event-triggering policy and multiple access protocol must be jointly analyzed in such a network. }\label{fig:NWcomparison}
\end{figure*}

The design of an event-based network must be accompanied by the selection of a suitable multiple access protocol, which determines the order of accessing the shared network. In a time-triggered network, transmission requests can be anticipated a priori, and a schedule can be drawn up to accommodate all the transmissions. In contrast, transmission requests cannot be anticipated in an event-based network. Thus, the access decisions must be taken at each sensor node, in a distributed manner. Furthermore, coordinating access decisions between nodes is not easy to accomplish on wireless networks. Thus, we choose to use a random access protocol; these protocols use a Contention Resolution Mechanism (CRM) to arbitrate access in a distributed, non-coordinated manner, between the nodes in the network. A protocol from the Carrier Sense Multiple Access (CSMA) family \cite{Rom1990}, called CSMA/CA (Collision Avoidance), is particularly well-suited to wireless networks and is used in Wifi \cite{ieee80211}, Zigbee \cite{zigbee} and WirelessHart \cite{WirelessHART2007}. In this paper, we use the $p$-persistent CSMA protocol, which provides an analytical approximation for the CRM in CSMA/CA \cite{Kleinrock1975}.

\subsection{Contribution}

An unavoidable consequence of distributed access decisions is packet collisions, which result when two or more nodes transmit at the same time. All the packets involved in a collision are lost, which can be detrimental to the performance of the closed-loop systems in the network. To minimize the impact of these collisions, one could adapt the event-triggering policy to the CRM response. However, such systems are inherently harder to analyze, as illustrated in Fig~\ref{fig:NWcomparison}. A typical network user generates traffic at a certain rate, and the network returns a probability of collision, which is a function of all the users' traffic rates. Thus, the user's rate and the performance of the multiple access protocol can be analyzed in isolation, in this case. In adaptive event-based networks, however, the traffic rate of each user is a function of the probability of collision of the network. Hence, a joint analysis of the event-triggering policy and the CRM is required.

Another consequence of random access is that network access for a node implies lack of access for all the other nodes in the network. Thus, the network access decisions are correlated, and for closed-loop systems, this correlation propagates to the system state. Closed-loop systems with exogenous noise processes become correlated due to their network interactions \cite{Cervin2008,Rabi2009}. Now, analyzing the resulting network is not a trivial task. To solve this problem, we derive inspiration from Bianchi's much-acclaimed analysis of the Distributed Coordination Function \cite{Bianchi2000} in IEEE~$802.11$. To counter a similar problem of network-induced correlation between traffic sources, Bianchi assumes that a node that is ready to transmit, sees a busy channel as a time-averaged, independent process. The independence aspect of this assumption restores a renewal property in our setup, enabling the use of a Markov model to represent the interactions in an event-based network. The time-average assumption permits a performance analysis in steady state. We verify these assumptions through simulations.

There are two main contributions of this paper. We present a joint analysis of the event-triggering policy and the CRM. The analysis is made possible by the use of Bianchi's assumption. In doing so, we also present a new configuration for the applicability of Bianchi's assumption. Our final contribution is the resulting network model; a Markov chain which describes the event-triggering policy and the multiple access protocol. With this model, we can view the event-triggering policy as a set of steady state probabilities. This model facilitates the design of a set of probabilities that ensure a system-level guarantee. In other words, the model and analysis presented in this paper can be used to design a network of event-based systems.

\subsection{Related Work}
Event-based systems were proposed as a means to reduce congestion in Networked Control Systems (NCS) \cite{Astrom1999,Yook2002,Otanez2002}. Early work showed that the same control performance can be achieved using fewer samples with event-based systems, for a single system \cite{Tomovic1966,Astrom1999}. Various event-triggering policies have been proposed for different problem formulations \cite{Rabi2006,Tabuada2007,Heemels2008,Henningsson2008}. However, the multiple access problem for event-based systems has not received as much attention. Much of the work focussing on the design of event-based systems for a shared network \cite{Wang2011,Molin2012} does not explicitly deal with the problem of multiple access. Others use protocols such as the CAN bus for wired networks \cite{Anta2009a}, or dynamic real-time scheduling for multiple tasks on a single processor \cite{Tabuada2007}. These protocols are not well-suited to wireless networks \cite{Akyildiz1999,Gummalla2000}.

There have been some attempts to analyze a network of event-based systems with random access. This includes a partial analysis of event-triggered nodes with CSMA/CA \cite{Cervin2008}, which highlighted the difficulties in analyzing such a network due to network-induced correlations. A more complete analysis with Aloha was presented in \cite{Rabi2009}, which assumed independent packet losses. A simple steady state model was presented in \cite{Henningsson2010}, but with an idealized multiple access protocol that results in no collisions. More recently, event-based systems which use Aloha and Slotted Aloha have been analyzed \cite{Blind2011a}, but with an event-triggering policy that is not adapted to the network. The work presented in this paper highlights the need for a joint analysis between the multiple access protocol and the event-triggering policy. An initial version of this work was presented in \cite{Ramesh2011b}.

\begin{figure*}[tb]
    \centering
    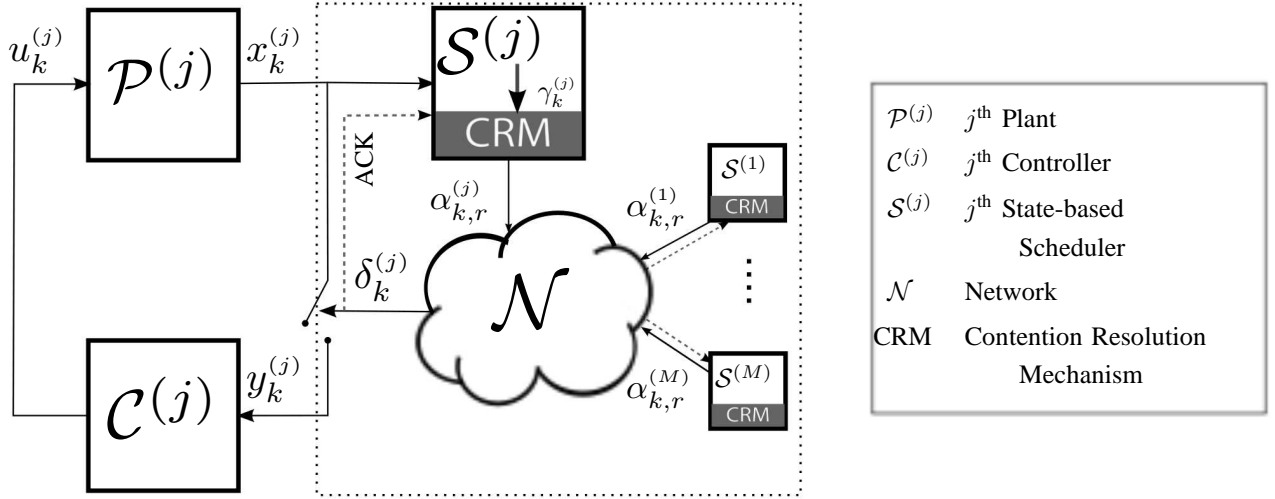
    \caption{A network of $M$ event-based systems using a CRM to access the network, where the $j^\textrm{th}$ control loop is illustrated with its dependency on the other control loops $i \in \{1,\dots,M\}$, $i \neq j$. The events are generated by the state-based schedulers. The other control loops are represented with their state-based schedulers alone.} \label{Fig:SystemModel}
\end{figure*}

\subsection{Outline}
The rest of the paper is organized as follows. We present the problem formulation in Section~\ref{S:Problem} and derive some important properties of the event-triggering policy, with no network traffic, in Section~\ref{S:ETprop}. We present the consequences of multiple access, and our solution using Bianchi's assumption in Section~\ref{S:MA_ET}. The Markov model describing the joint interactions, and the corresponding performance analysis are presented in this section. Finally, we present some simulation results in Section~\ref{S:Sims} and validate the assumptions of our model.

\section{Problem Formulation} \label{S:Problem}

\begin{figure*}[tb]
    \centering
    \def\svgwidth{\textwidth}
    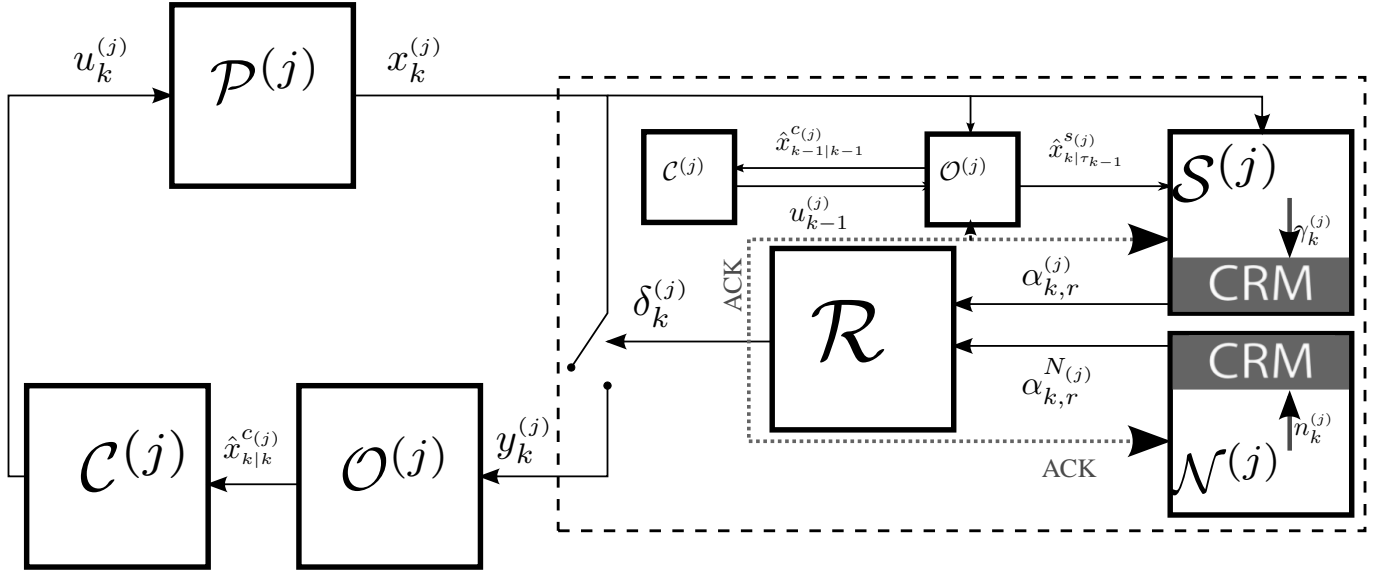
    \caption{A model of the network in Fig.~\ref{Fig:SystemModel}, from the perspective of a single event-based system. The event-triggering policy uses the prediction error to determine when to transmit. An explicit ACK is needed to track the estimation error at the sensor node. } \label{Fig:DualPred}
\end{figure*}

We consider a network of $M$ plants and controllers (indexed by $j \in \{1,\dots,M\}$), which communicate over a shared channel with an event-trigger in the loop, as shown in Fig.~\ref{Fig:SystemModel}. A model for the interactions between each event-based system and the network is depicted in Fig.~\ref{Fig:DualPred}. The blocks in this figure are explained below.

\noindent \textbf{Plant: } The plant $\mathcal{P}^{_{(j)}}$ has state dynamics given by
\begin{equation}
\label{Eq:StateSpace} x^{_{(j)}}_{k+1} = A_j x^{_{(j)}}_k + B_j u^{_{(j)}}_k + w^{_{(j)}}_k \; ,
\end{equation}
where $x_k \in \mathbb{R}^{n}$, $u_k \in \mathbb{R}^{m}$ and the initial state $x^{_{(j)}}_0$ and the process noise $w^{_{(j)}}_k$ are i.i.d. zero-mean Gaussians with covariance matrices $R^{_{(j)}}_0$ and $R^{_{(j)}}_w$, respectively. They are independent and uncorrelated to each other and to the initial states and process noises of other plants in the network. This discrete time model is defined with respect to a sampling period $T$ for each plant, and the sampling instants are generated by a synchronized network clock.

\noindent \textbf{State-based Scheduler: } There is a local scheduler $\mathcal{S}^{_{(j)}}$, situated in the sensor node, between the plant and the controller, which decides if the state $x^{_{(j)}}_k$ is to be ignored or selected for transmission. The scheduler output $\gamma^{_{(j)}}_k$ is correspondingly chosen from the set $\{0,1\}$, by the event-triggering policy $f^{_{(j)}}$, implemented within this block. The policy used in our setup is presented below, but motivated in Section~\ref{S:ETprop}. The scheduler output is given by
\begin{equation}
\gamma^{_{(j)}}_k = f^{_{(j)}}_k(x^{_{(j)}}_k-\hat{x}^{_{(j)}}_{F,k}) = \begin{cases}
1 & |x^{_{(j)}}_k-\hat{x}^{_{(j)}}_{F,k}|^2 > \Delta_{j}(m^{_{(j)}}_k) \; , \\
0 & \textrm{otherwise} \; ,
\end{cases} \label{Eq:InnoSched}
\end{equation}
where, $\Delta_{j}$ is the threshold, which typically depends on the memory index of the event-triggering policy $m^{_{(j)}}_{k}$. This index tracks the delay since the last received packet, $d^{_{(j)}}_{k-1}$, for delays smaller than the maximum memory index $F$, i.e., $m^{_{(j)}}_{k}= \min(d^{_{(j)}}_{k-1},F)$. In the above equation, $\hat{x}^{_{(j)}}_{F,k}$ plays the role of a memory-limited predicted estimate at the sensor node (\ref{Eq:PredEst}).

\noindent \textbf{Other Network Traffic}: The block $\mathcal{N}$ models a fictionalized source, representing traffic from all other event-based systems in the network. This traffic is represented by the network traffic index $n_k^{_{(j)}} \in \{0,1\}$.

\begin{figure}[tb]
    \centering
    \def\svgwidth{9cm}
    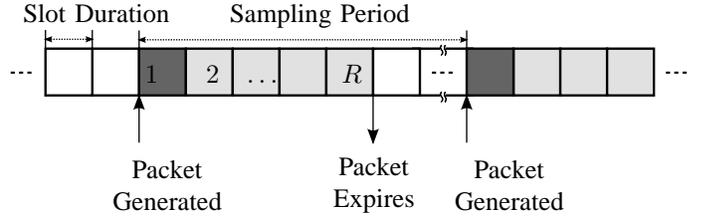
    \caption{The sampling period, corresponding to the system time scale, consists of many small slots, corresponding to the CRM time scale. For each packet generated by the event-based system, the CRM attempts $R$ retransmissions before it declares a packet loss due to congestion. This feature increases the reliability of the multiple access protocol.} \label{Fig:TimeScales}
\end{figure}

\noindent \textbf{CRM}: The multiple access protocol implements a CRM in each sensor node, which resolves contention between simultaneous channel access requests in a distributed manner. We consider the $p$-persistent CSMA mechanism, with $R$ retransmissions. The retransmissions occur in the CRM time scale, which is much finer in resolution than the system time scale, as indicated in Fig.~\ref{Fig:TimeScales}. The time scales are assumed to be separated, i.e., all retransmissions corresponding to a single event are completed before the next sampling period. The access indicator for each retransmission $r=\{1,\dots,R\}$ is denoted $\alpha^{_{(j)}}_{k,r} \in \{0,1\}$ at time $k$. The persistence probability in the $r^{\textrm{th}}$ retransmission attempt is defined as $\Pr(\alpha^{_{(j)}}_{k,r} = 1 | \gamma^{_{(j)}}_{k} = 1)$, and denoted by $p_{\alpha,r}$ for brevity.

\noindent \textbf{Channel Access Indicator}: The channel access indicator $\delta_k^{_{(j)}} \in \{0,1\}$ denotes transmission failure or success, respectively, after $R$ retransmission attempts. A transmission is successful if there is only one system that attempts to access the channel in that CRM slot, as  given by
\begin{equation}
\label{Eq:Delta}
\delta^{_{(j)}}_k = \bigvee_{r=1}^{R} \left [ \alpha^{_{(j)}}_{k,r} \cdot (1-\alpha^{N_{(j)}}_{k,r})  \right ] \; ,
\end{equation}
and then, $\sum_{j=1}^{M} \delta^{_{(j)}}_k \le R$. Thus, the number of retransmissions, $R$, also determines the maximum number of transmissions supported by the network protocol, for every system sampling instant.

\noindent \textbf{Observer:} The observer $\mathcal{O}^{_{(j)}}$ receives $y^{_{(j)}}_k$, given by
\begin{equation} \label{Eq:NLMeas}
y^{_{(j)}}_k = \begin{cases}
x^{_{(j)}}_k & \delta^{_{(j)}}_k = 1 \; , \\
\varepsilon & \textrm{otherwise} \; ,
\end{cases}
\end{equation}
where $\varepsilon$ denotes a packet erasure when there is no event. The estimate is computed as
\begin{equation} \label{Eq:EstO}
\hat{x}^{c_{(j)}}_{^{k|k}} = \begin{cases}
x^{_{(j)}}_k & \delta^{_{(j)}}_k = 1 \; , \\
A_j \hat{x}^{c_{(j)}}_{^{k-1|k-1}} + B_j u^{_{(j)}}_{k-1}  & \textrm{otherwise} \; ,
\end{cases}
\end{equation}
with $\hat{x}^{c_{(j)}}_{^{-1|-1}} = 0$. We define the corresponding estimation error as $\tilde{x}^{{(j)}}_{k} = x^{_{(j)}}_k - \hat{x}^{c_{(j)}}_{^{k|k}}$. A copy of this observer is used at the sensor node to facilitate the event-triggering policy by generating the predicted estimate $\hat{x}^{s_{(j)}}_{k|\tau_{k-1}} = A_j^{(k-\tau_{k-1})} x^{_{(j)}}_{\tau_{k-1}} + \sum_{l=\tau_{k-1}}^{k-1} A_j^{(k-l-1)} B_j u^{_{(j)}}_{l}$. This is used to generate the memory-limited predicted estimate $\hat{x}^{_{(j)}}_{F,k}$ in (\ref{Eq:InnoSched}), which is given by
\begin{equation}
\hat{x}^{_{(j)}}_{F,k} = \begin{cases}
\hat{x}^{s_{(j)}}_{k|\tau_{k-1}} & d^{_{(j)}}_{k-1} < F \; , \\
\hat{x}^{s_{(j)}}_{k|k-F} & \textrm{otherwise} \; .
\end{cases} \label{Eq:PredEst}
\end{equation}
Thus, $\hat{x}^{_{(j)}}_{F,k}$ is given by the predicted estimate $\hat{x}^{s_{(j)}}_{k|\tau_{k-1}}$ when the delay is less than the memory $F$ of the scheduler. When the delay exceeds this value, $\hat{x}^{_{(j)}}_{F,k}$ takes the value $\hat{x}^{s_{(j)}}_{k|k-F}$, which is generated by assuming knowledge of $x^{_{(j)}}_{k-F}$, in place of $x^{_{(j)}}_{\tau_{k-1}}$.

\noindent \textbf{Controller:} The controller $\mathcal{C}^{_{(j)}}$ generates an appropriate control signal, such as
\begin{equation}
u^{_{(j)}}_k = -L^{_{(j)}}_k  \hat{x}^{c_{(j)}}_{^{k|k}} \;  , \label{Eq:Controller}
\end{equation}
where $L^{_{(j)}}_k$ is selected to minimize an appropriate cost function, such as the linear quadratic Gaussian cost.

\begin{figure}[tb]
    \centering
    \def\svgwidth{9cm}
    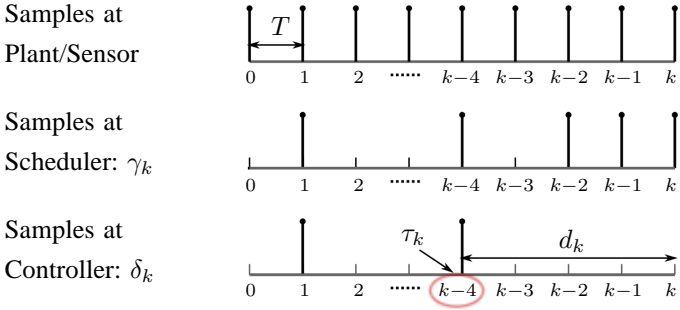
    \caption{An illustration of the delay since the last received packet $(d_k)$ and the index of the last received packet $(\tau_k)$. The topmost sample trace indicates the discrete sampling instants, the middle one indicates the events selected at the state-based scheduler, and the bottom one indicates the events that successfully reach the controller. } \label{Fig:TimeLines}
\end{figure}

We are interested in analyzing the joint performance of the event-trigger and CRM in this network, in steady state. To do so, we define two metrics that characterize the network performance.
\begin{definition}[Steady-state Delay Distribution]
The delay since the last received packet is given by $d^{_{(j)}}_k = k - \tau^{_{(j)}}_k$, where $\tau^{_{(j)}}_k$ is the time index of the last received packet, as illustrated in Fig.~\ref{Fig:TimeLines}. To avoid notational overhead, we skip the index $j$ for $\tau_k$ and $d_k$, when the context is clear. The time index of the last received packet is defined as $\tau^{_{(j)}}_k = \max\{t: \delta^{_{(j)}}_t=1\}$, for $-1 \le t \le k$ and $\delta^{_{(j)}}_{-1}=1$. Note that $-1 \le \tau^{_{(j)}}_k \le k$. Then, the steady-state delay distribution is defined as $\Pr^{_{(j)}}_{d}(\zeta) \triangleq \lim_{k \rightarrow \infty} \Pr(d^{_{(j)}}_k = \zeta)$, for $\zeta \in \mathbb{Z}$.
\end{definition}

\begin{definition}[Steady-state Reliability]
Recall that $\delta^{_{(j)}}_k$ is the channel access indicator. The steady state probability of a successful transmission as a consequence of the joint actions of the event-trigger and CRM is defined as $p^{_{(j)}}_{\delta} \triangleq \lim_{k \rightarrow \infty} \Pr(\delta^{_{(j)}}_k = 1)$. This indicates the network reliability on the sensing link for a given closed-loop system.
\end{definition}

The above information is a prerequisite for any design methodology that seeks to achieve a certain network or system guarantee.

\subsection{Motivating Example}
\begin{figure}[tb]
\begin{center}
\includegraphics*[scale=0.2,viewport=75 75 1400 525]{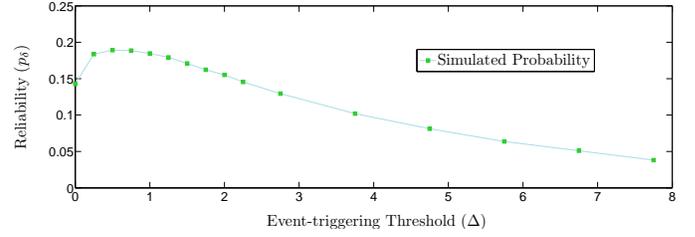}
\caption{A plot of the simulated values of the reliability versus the scheduler threshold. This plot clearly shows that the above relationship cannot be approximated by an i.i.d. loss process or any other such simplistic modelling technique. } \label{Fig:pDeltaVSepsilonTeaser}
\end{center}
\vspace{-5mm}
\end{figure}

Before we delve into the main results, we present an example of a network of systems, and examine a performance analysis curve for this example obtained using Monte-Carlo simulations. With this example, we wish to motivate the methods used in the rest of this paper.
\begin{example}[Network and Experiment Setup] \label{Ex:Setup}
We consider a homogenous network of $M=10$ nodes, with $R=5$ retransmissions in the CRM. The dynamics of the plants are given by \eqref{Eq:StateSpace} for $x_k \in \mathbb{R}$ and $w_k \sim \mathcal{N}(0,1)$. The plants are identical with state transition matrix $A = 1$ and control matrix $B = 1$. We use a state-based scheduler (\ref{Eq:InnoSched}) with the event-triggering policy $| x_k - \hat{x}_{F,k} | ^2 > \Delta$, where $\Delta$ is a constant scheduler threshold and $\hat{x}_{F,k}$ denotes the memory-limited predicted estimate (\ref{Eq:PredEst}) at time $k$. When the delay exceeds the memory of the policy $F$, the value $x_{k-F}$ is assumed to be the last successfully received value while computing $\hat{x}_{F,k}$, thus limiting the memory of the event-triggering policy. To realize a scheduler such as this, we implement the dual predictor architecture presented in Fig.~\ref{Fig:DualPred}. The CRM used to arbitrate access is the $p$-persistent CSMA protocol, and $p_\alpha = 0.2$ for all $5$ retransmission stages of the CRM.

A plot of the simulated values of reliability $p_\delta$ versus the scheduler threshold $\Delta \in (0,8)$ is shown in Fig.~\ref{Fig:pDeltaVSepsilonTeaser}. This plot has been obtained using Monte-Carlo simulations. The non-linear relationship depicted in the plot is not surprising, considering that a given scheduler threshold translates to a certain traffic rate depending on the distribution of the estimation error. However, it is important to note that the distribution of the estimation error with delay evolves based on the probability of a successful transmission, as a consequence of the adaptation illustrated in Fig.~\ref{fig:NWcomparison}. Thus, it is apparent from Fig.~\ref{Fig:pDeltaVSepsilonTeaser} that there is no simple loss process that captures the interaction of a single system with the rest of the network.
\end{example}

We return to this example in Section~\ref{S:Sims}, where we comment on the non-monotonic relationship obtained from simulations.

\section{The Event-triggering Policy} \label{S:ETprop}

We examine our event-triggering policy, to understand what it does for a single closed-loop system without other network traffic. We show that this policy adapts to the estimation error, or a part of it, when its memory is constrained. The renewal property of the estimation error is used to construct a Markov model that represents the functioning of the event-triggering policy. Since we only consider a single closed-loop system in this section, we drop the index $j$. The lack of other network traffic implies that $n_k \equiv 0$ for all $k \ge 0$. Now, there is no need for a multiple access protocol and every event results in a successful transmission, i.e., $\delta_k = \gamma_k$.

\subsection{Properties of the Event-triggering Policy}
We begin by motivating our selection of the event-triggering policy in (\ref{Eq:InnoSched}). The events generated by our policy are not influenced by the past applied controls, resulting in a structural separation between the state-based scheduler, observer and controller, as shown in \cite{Ramesh2013}. Thus, the role of the controller is limited to regulating the estimate (\ref{Eq:EstO}), and the role of the state-based scheduler is limited to reducing the estimation error. Accordingly, the policy defined in (\ref{Eq:InnoSched}) adapts to the estimation error across the network; the input to this policy is the estimation error, for delays not exceeding $F$, or a related quantity, when the delay is $F$ or larger. This can be seen from
\begin{align}
|x_k-\hat{x}_{F,k}|^2 &= \begin{cases}
\left| x_k-\hat{x}^{s}_{k|\tau_{k-1}} \right|^2 & d_{k-1} < F \; , \\
\left| x_k-\hat{x}^{s}_{k|k-F} \right|^2 & \textrm{otherwise}
\end{cases} \notag \\
&= \begin{cases}
\left| \sum_{l=\tau_{k-1}}^{k-1} A^{(k-l-1)} w_{l} \right|^2 & d_{k-1} < F \; , \\
\left| \sum_{l=k-F}^{k-1} A^{(k-l-1)} w_{l} \right|^2 & \textrm{otherwise} \; .
\end{cases} \label{Eq:ETesterrip}
\end{align}
Note that the estimation error for $d_{k-1} \ge F$ is given by $\sum_{l=\tau_{k-1}}^{k-1} A^{(k-l-1)} w_{l}$. However, the value used in its place in the event-triggering policy is obtained by assuming that $x_{k-F}$ was successfully transmitted, i.e., $\tau_{k-1} = k-F$. Thus, the statistical properties of the inputs to the above event-triggering policy vary with delay for $d_{k-1} < F$, but remain constant for $d_{k-1} \ge F$. Hence, we limit the memory of our adaptive policy to $F$.

Following a successful transmission, the estimation error is reset to zero at the observer, and this leads to some desirable properties for our policy, discussed below. For a sequence $a_k$, the notation $\vect{a}_{t_0}^{t_f}$ is used to denote the set $\{a_{t_0},\dots,a_{t_f}\}$.
\begin{lemma} \label{Lemma:Markovian_EstErr}
For a single system given by (\ref{Eq:StateSpace})--(\ref{Eq:InnoSched}), (\ref{Eq:EstO})--(\ref{Eq:Controller}), with $\delta_k = \gamma_k$, $e_k = \vect{\tilde{x}}_{^{\tau_k}}^{_{k}}$ is a Markovian representation for the estimation error at the observer, $\tilde{x}_k$. In other words,
\begin{equation} \label{Eq:Markovian_EstErr}
\Pr(e_k|\vect{e}_{^{0}}^{_{k-1}}) = \Pr(e_k|e_{k-1}) \; .
\end{equation}
\end{lemma}
\begin{IEEEproof}
At any time $k$, $\tau_k$ represents the time index corresponding to the last received packet. Then, $\tilde{x}_{\tau_k} = 0$, as the state $x_{\tau_k}$ is received by the observer. At time $\tau_k+1$, the estimation error corresponds to the process noise $w_{\tau_k}$. The process noise is i.i.d., and hence, independent of the state at $\tau_k$ or prior to it. This is also true for any future estimation error. Thus, we have
\begin{equation*}
\Pr(\tilde{x}_{k}|\vect{y}_{^{0}}^{_{k}},\vect{\delta}_{^{0}}^{_{k}}) = \Pr(\tilde{x}_{k}|\vect{y}_{^{\tau_k}}^{_{k}},d_k) \; .
\end{equation*}
The delay since the last transmission, along with the measurement values since the last transmission form a sufficient statistic for the estimation error. Using this, and the relationship $\tau_k = \tau_{k-1}$ when $\delta_k = 0$, we obtain (\ref{Eq:Markovian_EstErr}).
\end{IEEEproof}

\begin{corollary} \label{Corollary:ETrenewal}
For a single system given by (\ref{Eq:StateSpace})--(\ref{Eq:InnoSched}), (\ref{Eq:EstO})--(\ref{Eq:Controller}), with $\delta_k = \gamma_k$, the inter-arrival times at the controller are independent.
\end{corollary}
\begin{IEEEproof}
The inter-arrival times are given by $t_{i+1} - t_i$, where $\{t: d_{t}=0\}$ denotes the packet reception instants. Following the successful reception of a packet at time $t_i$, the future estimation error is independent of $\tilde{x}_{^{t_{i}|t_{i-1}}}$. For the event-triggering policy in (\ref{Eq:InnoSched}), the estimation error $\tilde{x}_{^{t_{i}+s|t_{i}}}$, for $s>1$, determines $t_{i+1}$. Thus, $t_{i+1} - t_i$ is independent of $t_i - t_{i-1}$.
\end{IEEEproof}
From Corollary~\ref{Corollary:ETrenewal}, we can thus conclude that the event-triggering policy in (\ref{Eq:InnoSched}) results in a traffic source that is a renewal process.

\subsection{Markov Chain Representation} \label{S:MCnotraffic}

\begin{figure*}[tb]
    \centering
    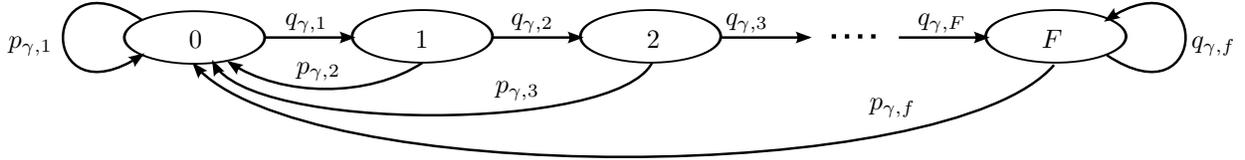
    \caption{A Markov chain model representing the event-triggering policy in (\ref{Eq:InnoSched}), when there is no exogenous network traffic. The estimation error grows with delay, resulting in different probabilities for events and non-events until $F$. } \label{Fig:MCet}
\end{figure*}

Using Lemma~\ref{Lemma:Markovian_EstErr}, we construct a Markov chain to represent the event-triggering policy, as shown in Fig.~\ref{Fig:MCet}. The state indices $m=\{0,\dots,F\}$ represent the memory of the event-triggering policy in (\ref{Eq:InnoSched}). A return to the state $m=0$ denotes a successful transmission, when the estimation error is reset to zero. From here on, the number of terms contributing to the input of the event-triggering policy continue to grow, as can be seen from the expression $\sum_{l=\tau_{k-1}}^{k-1} A^{(k-l-1)} w_{l}$. For $m<F$, we see two transitions out of every state; one to the next state $m+1$ indicating a non-transmission, and the other to $0$ indicating a successful transmission. The corresponding probabilities are denoted $q_{\gamma,m+1}$ and $p_{\gamma,m+1}$, respectively, and defined as
\begin{equation} \label{Eq:TransProb_pgm}
\begin{aligned}
p_{\gamma,m+1} &= \Pr\bigg( \bigg| \sum_{l=k-(m+1)}^{k-1} A^{(k-l-1)} w_{l} \bigg|^2 > \Delta(m_{k}) \\
& \quad \quad \quad \quad \quad \quad \quad \bigg| d_{k-1}=m_k \bigg) \; , \quad m<F \; , \\
q_{\gamma,m+1} &= 1- p_{\gamma,m+1} \; .
\end{aligned}
\end{equation}
For $m=F$, there are two transitions again, but a non-transmission returns to the same state, with probability $q_{\gamma,f}=1-p_{\gamma,f}$. The probability $p_{\gamma,f}$ is defined as in (\ref{Eq:TransProb_pgm}), with $m = F-1$.

\begin{remark}[Event Probabilities:]
The probabilities of events and non-events in (\ref{Eq:TransProb_pgm}) can be computed given the event thresholds in (\ref{Eq:InnoSched}), though this computation is not trivial as the estimation error does not have a Gaussian distribution. However, an event-triggered policy can be specified both in terms of event-thresholds or event-probabilities. For the rest of this paper, we assume that the event-triggering policy in (\ref{Eq:InnoSched}) is specified in terms of event probabilities, rather than event thresholds. The conversion from event probabilities to event thresholds becomes relevant when the event-triggering policy must be implemented, and we deal with that in Section~\ref{S:Sims}.
\end{remark}

\begin{remark}[Effect of independent Packet Losses:]
It is straightforward to extend the above model to include independent packet losses, which occur with probability $p_L$. A simple change of variables, with $p^{L}_{\gamma,m} = p_{\gamma,m} \cdot (1-p_L)$ in place of $p_{\gamma,m}$ and $q^L_{\gamma,m} = 1 - p^L_{\gamma,m}$ in place of $q_{\gamma,m}$ gives us our modified Markov chain. This is because the future estimation error and events remain independent of the past, after a transmission, and hence the statement of Lemma~\ref{Lemma:Markovian_EstErr} continues to hold when there are packet losses.
\end{remark}

\section{The Multiple Access Event-triggered Problem} \label{S:MA_ET}

We now look at what happens when there are many event-based systems in the same network, i.e., $n_k \neq 0$. In this case, the CRM plays an important role as $\delta_k \neq \gamma_k$. We first examine the consequences of introducing the CRM in the network. Then, we use introduce Bianchi's assumption, and use this to construct a Markov model to represent the dynamics of the event-triggering policy and CRM.

\subsection{Consequences of the CRM} \label{S:MAconsequences}
The CRM impacts the network in two ways. The first consequence is that the event-triggered policy must be jointly analyzed with the CRM, as has already been illustrated in Fig.~\ref{fig:NWcomparison}. The other consequence is the correlation introduced between the various systems due to network interactions. We state and prove this below. 

\begin{lemma} \label{Lemma:CorrelatedNW}
For the system described by (\ref{Eq:StateSpace})--(\ref{Eq:Controller}), the estimation errors corresponding to different plants in the network are correlated, i.e., 
\begin{equation} \label{Eq:CorrelatedNW}
\Pr(\tilde{x}^{_{(1)}}_{k},\dots,\tilde{x}^{_{(M)}}_{k}) \neq \prod_{j=1}^M \Pr(\tilde{x}^{_{(j)}}_{k} ) \; .
\end{equation}
\end{lemma}
\begin{IEEEproof}
A network that supports $R$ retransmissions must satisfy the constraint $\sum_{j=1}^{M} \delta^{_{(j)}}_k \le R$. Due to this, and the definition of $\delta_k$ in (\ref{Eq:Delta}), the probability of a successful transmission depends on the probabilities of all the events in the network at time $k$. This can be expressed as
\begin{align} \label{Eq:corrPrdelta}
\Pr(\delta^{_{(j)}}_k | \gamma^{_{(1)}}_{k},\dots,\gamma^{_{(j)}}_{k}=1,\dots,\gamma^{_{(M)}}_{k}) &\neq \Pr(\delta^{_{(j)}}_k | \gamma^{_{(j)}}_{k}=1) \; .
\end{align}
The estimate (\ref{Eq:EstO}) and the corresponding estimation error $\tilde{x}^{_{(j)}}_{k}$ are determined by $\delta_k$. Hence, the estimation error is correlated to all the events at time $k$. This is true for all the plants in the network, and thus, they become correlated to one another as indicated in (\ref{Eq:CorrelatedNW}).
\end{IEEEproof}

The above result reaffirms that the CRM introduces correlations between different event-based systems, as has been noted earlier in \cite{Cervin2008,Rabi2009}. The correlation between the estimation errors leads to correlation in the states, prediction errors and future scheduler outputs. An example of a scenario that might arise due to the above result is as follows; a large estimation error in a system that does not get to transmit, perhaps due to random access or collisions, might result in increased congestion for the entire network due to persistent events from this system. This in turn might cause the estimation error to grow in other systems, and lead to further congestion. Hence, the properties in Lemma~\ref{Lemma:Markovian_EstErr} and Corollary~\ref{Corollary:ETrenewal} do not hold for the systems in such a network, as formally proven below.

\begin{lemma} \label{Lemma:LostProp}
For the system described in (\ref{Eq:StateSpace})--(\ref{Eq:Controller}), $e^{_{(j)}}_k = \{\vect{\tilde{x}}^{_{(j)}}\}_{^{\tau_k}}^{_{k}}$ is not a Markovian process. Consequently, the inter-arrival times at the observer are not independent.
\end{lemma}
\begin{IEEEproof}
The Markovian properties of $e^{_{(j)}}_k$ in Lemma~\ref{Lemma:Markovian_EstErr} followed from the independence of the estimation error, following a transmission, from its past. This is no longer true when there are interactions through the CRM. To see this, let us examine the prediction and estimation error following a transmission instant, $\tau^{_{(j)}}_k$, for the $j^{\textrm{th}}$ plant and for some $k \ge 0$. The prediction error is given by $\tilde{x}^{_{(j)}}_{\tau_k+1|\tau_k} = w^{_{(j)}}_{\tau_k}$, and it is independent of the estimation error prior to $\tau^{_{(j)}}_k$ due to the independence of the process noise $w^{_{(j)}}_{\tau_k}$. Thus, we have
\begin{equation*}
\Pr(\tilde{x}^{_{(j)}}_{\tau_k+1|\tau_k} | \tilde{x}^{_{(j)}}_{\tau_k|\tau_k}) = \Pr(\tilde{x}^{_{(j)}}_{\tau_k+1|\tau_k}) \; .
\end{equation*}

Consequently, $\gamma^{_{(j)}}_{\tau_{k}+1}$ is independent of all the other scheduler outputs. However, $\delta^{_{(j)}}_{\tau_{k}+1}$ is still determined by all the scheduler outputs at time $\tau^{_{(j)}}_k + 1$, as shown in (\ref{Eq:corrPrdelta}). Thus, the estimation error $\tilde{x}^{_{(j)}}_{\tau_k+1|\tau_k + 1}$ is correlated with the estimation errors from all the other plants in the network, as shown in Lemma~\ref{Lemma:CorrelatedNW}, some of which may be correlated with the estimation error of plant $j$ prior to $\tau^{_{(j)}}_k + 1$. Thus, the network interaction reintroduces a correlation with its past, and the estimation error following a reception instant is not independent of its past. In other words,
\begin{equation*}
\Pr(\tilde{x}^{_{(j)}}_{\tau_k+1|\tau_k+1} | \tilde{x}^{_{(j)}}_{\tau_k|\tau_k}) \neq \Pr(\tilde{x}^{_{(j)}}_{\tau_k+1|\tau_k+1}) \; .
\end{equation*}
Consequently, $e_k$ is not Markovian. The lack of independence implies that arrival times are also correlated in this setup.
\end{IEEEproof}

A successful transmission for a node in a congested network need not reduce congestion for the other nodes in the network. The event-backlog may take a few sampling periods to dissipate. In the meanwhile, new events from the successful nodes will continue to see traffic conditions similar to those encountered by previous events from these nodes. Thus, the independence of the estimation error following a successful transmission is lost due to these network interactions. Analyzing the joint performance of the event-triggering policy and CRM is a challenging task due to the correlations in the network.

\subsection{Bianchi's Assumption} \label{S:BianchiAssumption}

We now use an assumption from Bianchi's seminal paper \cite{Bianchi2000} that simplifies the network interactions. While presenting this assumption and utilizing it to construct a model, we consider the simplest setup in the multiple access network, which corresponds to the case when the CRM permits no retransmissions, i.e., $R=1$. Accordingly, we denote the CRM access indicator $\alpha_{k,1}$ simply as $\alpha_k$, with corresponding probability $p_{\alpha}$. These results can be extended to include multiple retransmissions, which we discuss in Section~\ref{S:Extn}.

\begin{assumption} \label{Assumption:Bianchi}
For the systems described in (\ref{Eq:StateSpace})--(\ref{Eq:Controller}), the conditional probability of a busy channel for a node that attempts to transmit in steady state, is given by an independent probability $p$ for each node. Thus,
\begin{equation} \label{Eq:Bianchi}
\lim_{k \to \infty} \Pr(\delta^{_{(j)}}_k=0|\gamma^{_{(j)}}_k=1,\alpha^{_{(j)}}_{k}=1) = p^{_{(j)}} \; ,
\end{equation}
for all $j \in \{1,\dots,M\}$.
\end{assumption}

Simulations in Section~\ref{S:Sims} validate this assumption as a reasonable one to make for our problem setup. There are two aspects to this assumption; Firstly, (\ref{Eq:Bianchi}) removes the correlation of the channel access indicator $\delta^{_{(j)}}_{k}$ with the scheduler outputs of all the other plants in the network, which was shown in (\ref{Eq:corrPrdelta}). Secondly, notice that $p^{_{(j)}}$ is not indexed by $k$; it is a time-average, and results in a steady state analysis, as we show in the rest of the paper. Now, we use the independence aspect to extend the desirable properties of Lemma~\ref{Lemma:Markovian_EstErr} and Corollary~\ref{Corollary:ETrenewal}, to systems in the multiple access network.

\begin{theorem} \label{Theorem:Markovian_EstErr}
For the systems described in (\ref{Eq:StateSpace})--(\ref{Eq:Controller}), with Assumption~\ref{Assumption:Bianchi}, $e^{_{(j)}}_k = \{\vect{\tilde{x}}^{_{(j)}}\}_{^{\tau_k}}^{_{k}}$ is a Markovian representation for the steady state estimation error at the observer, $\tilde{x}^{_{(j)}}_{k}$, for all $j \in \{1,\dots,M\}$. In other words,
\begin{equation} \label{Eq:Markovian_EstErrMultiple}
\lim_{k \to \infty} \Pr(e^{_{(j)}}_k|\{\vect{e}^{_{(j)}}\}_{^{0}}^{_{k-1}}) = \lim_{k \to \infty} \Pr(e^{_{(j)}}_k|e^{_{(j)}}_{k-1}) \; .
\end{equation}
Consequently, the inter-arrival times at the observer for each plant is independent.
\end{theorem}
\begin{IEEEproof}
In Corollary~\ref{Lemma:LostProp}, we showed that the estimation error following a packet reception instant $\tau^{_{(j)}}_{k}+1$, for some $j \in \{1,\dots,M\}$, is not independent of its past due to the correlation introduced by $\delta^{_{(j)}}_{\tau_{k}+1}$. Re-examining (\ref{Eq:corrPrdelta}), with Assumption~\ref{Assumption:Bianchi}, we now get
\begin{align*}
\Pr(\delta^{_{(j)}}_k &| \gamma^{_{(1)}}_{k},\dots,\gamma^{_{(j)}}_{k}=1,\dots,\gamma^{_{(M)}}_{k}) \\ 
&= \sum_{\alpha^{_{(j)}}_k \in \{0,1\}} \Pr(\delta^{_{(j)}}_k | \gamma^{_{(j)}}_{k}=1, \alpha^{_{(j)}}_k) \cdot \Pr(\alpha^{_{(j)}}_k | \gamma^{_{(j)}}_k=1) \; ,
\end{align*}
which implies that
\begin{align*}
\lim_{k \to \infty} \Pr(\delta^{_{(j)}}_k = 0 | \gamma^{_{(1)}}_{k},\dots,\gamma^{_{(j)}}_{k}=1,\dots,\gamma^{_{(M)}}_{k}) &= p^{_{(j)}} \cdot p_{\alpha} \\
& \quad \quad + 1 \cdot q_{\alpha} \; , \\
\lim_{k \to \infty} \Pr(\delta^{_{(j)}}_k = 1 | \gamma^{_{(1)}}_{k},\dots,\gamma^{_{(j)}}_{k}=1,\dots,\gamma^{_{(M)}}_{k}) &= q^{_{(j)}} \cdot p_{\alpha} \; .
\end{align*}
Thus, the dependence on the other scheduler outputs vanishes due to Assumption~\ref{Assumption:Bianchi}. Now, the estimation error remains independent of its past, i.e.,
\begin{equation*}
\lim_{k \to \infty} \Pr(\tilde{x}^{_{(j)}}_{\tau_k+1|\tau_k+1} | \tilde{x}^{_{(j)}}_{\tau_k|\tau_k}) = \lim_{k \to \infty} \Pr(\tilde{x}^{_{(j)}}_{\tau_k+1|\tau_k+1}) \; .
\end{equation*}
Thus, using the same arguments as in the proof of Lemma~\ref{Lemma:Markovian_EstErr}, we can establish the Markovian property of $e^{_{(j)}}_k$ in (\ref{Eq:Markovian_EstErrMultiple}). Consequently, the inter-arrival times are independent.
\end{IEEEproof}

Bianchi's assumption has converted the traffic source corresponding to the event-triggering policy and the CRM, into a renewal process. Now, analyzing the performance of this network is straightforward.

\subsection{Markov Chain Representation} \label{S:MCtraffic}

\begin{figure*}[tb]
\begin{center}
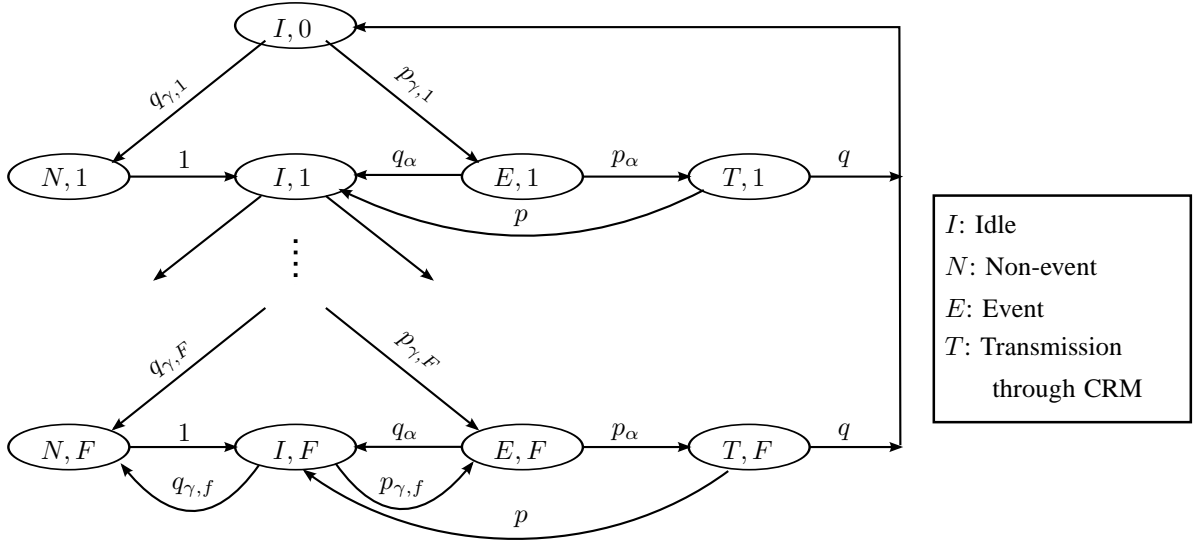
\caption{A Markov chain representation for the event-triggering policy in (\ref{Eq:InnoSched}) and a simple CRM with no retransmissions. The variable $F$ denotes the memory limit of the transmission history used by the scheduler.} \label{Fig:MCetCRM}
\end{center}
\vspace{-5mm}
\end{figure*}

We use Bianchi's assumption to construct a Markov model of the event-triggering policy and CRM. The presentation in this section corresponds to a single system in the network, and thus, we skip the index $(j)$.

In the Markov chain in Fig.~\ref{Fig:MCetCRM}, we assign two indices, $(S,m)$, to each state and denote the probability of being in the state as $p_{_{(S,m)}}$. The index $m$ represents the steady state memory of the scheduler and is given by $\min(d_k,F)$. The states $(S,m)$ and $(S,m+1)$ are one sampling period away from each other. The index $S$ represents the four states a packet can be in during a sampling period. These are
\begin{enumerate}
\item $S=I$ [Idle State]: For $m>0$, non-events and unsuccessful events return to this state before the next sampling instant. The initial state $(I,0)$ indicates the idle state before the next sampling instant following a successful transmission.
\item $S=N$ [Non-event State]: This state is reached when the scheduler output $\gamma_k=0$. A transition out of this state occurs instantaneously, and always to the idle state to wait for the next sampling instant.
\item $S=E$ [Event State]: This state is reached when $\gamma_k=1$. A transition out of this state occurs to the transmission state or the idle state, depending on the CRM access indicator $\alpha_k$. When $\alpha_k=0$, the event is discarded and the system moves to the idle state to wait for the next sampling instant.
\item $S=T$ [Transmission in CRM State]: The CRM's inclusion can be seen directly in this state; it is reached only when the CRM permits channel access, or when $\alpha_k=1$. Note that only a node in state $T$ actually attempts a transmission. A transition out of this state occurs instantaneously, with two possibilities: transmission success or failure.
\end{enumerate}

The transition probabilities in Fig.~\ref{Fig:MCetCRM} are explained below:
\begin{itemize}
\item $p_{\gamma,m}$ and $q_{\gamma,m}$ denote the probability of an event and non-event respectively, and are defined in (\ref{Eq:TransProb_pgm}).
\item $p_{\gamma,f}$ and $q_{\gamma,f}$ denote the probability of an event and non-event, respectively, when $d_{k-1} \ge F$.
\item $p_\alpha$ denotes the probability of accessing the channel through the CRM. Conversely, $q_\alpha = 1-p_\alpha$ represents the probability of discarding an event.
\item $p$ denotes the conditional probability of a busy channel, as defined in (\ref{Eq:Bianchi}). A successful transmission occurs with probability $q$.
\end{itemize}

Note that the Markov chain in Fig.~\ref{Fig:MCetCRM} represents the event-triggering and CRM of one system in the network. Thus, each system has its own such Markov chain, and these interact to produce the busy channel process in (\ref{Eq:Bianchi}).

\section{Steady State Performance Analysis} \label{S:SSanal}

In this section, we use the time-averaged aspect of Bianchi's assumption along with the Markov chain in Fig.~\ref{Fig:MCetCRM} to derive a steady state analysis. We also present extensions to more advanced network settings.

\subsection{Steady State Performance}
\begin{theorem} \label{Theorem:SteadyStateAnalysis}
For a system described by (\ref{Eq:StateSpace})--(\ref{Eq:Controller}), with Assumption~\ref{Assumption:Bianchi}, the network reliability is given by
\begin{equation} \label{Eq:PrDelta1}
p^{_{(j)}}_{\delta} = (1-p^{_{(j)}}) \cdot p^{_{(j)}}_{\mathrm{tx}} \; ,
\end{equation}
where, $\smash{p^{_{(j)}}}$ is the conditional probability of a busy channel for nodes attempting to transmit as defined in (\ref{Eq:Bianchi}), and $\smash{p^{_{(j)}}_{\mathrm{tx}} = \sum_{m=1}^F p^{_{(j)}}_{_{(T,m)}}}$ is the steady state probability that a node attempts to transmit, or is in any of the $(T,m)$ states.
\end{theorem}
\begin{IEEEproof}
We begin by evaluating the probabilities $p_{_{(S,m)}}$, in steady state, using the transition probabilities defined above. Then, we describe the interaction between the Markov models (Fig.~\ref{Fig:MCetCRM}) corresponding to each of the systems in the network, to find an expression for the probability of a successful transmission.

The state $(I,m)$, for $m>0$, is always reached unless there is a successful transmission. The probability of a successful transmission in the $m^{th}$ stage is given by $p_{\gamma,m} p_\alpha q$. Thus, we obtain the recursive expression
\begin{align}
p_{_{(I,m)}} &= (1 - p_{\gamma,m} p_\alpha q) p_{_{(I,m-1)}} \; , \quad m = 1,\dots,F-1 \; , \notag \\
p_{_{(I,F)}} &= \frac{1 - p_{\gamma,F} p_\alpha q}{p_{\gamma,f} p_\alpha q} p_{_{(I,F-1)}} \; . \label{Eq:pIdleState}
\end{align}
In the final stage, $(I,F)$ can be reached from state $(I,F-1)$ and from state $(I,F)$ itself, which gives us the above equation. Also, at any sampling instant, a node must be in any of the $(I,m)$ states. Thus, we have
\begin{equation}
\label{Eq:sumb0}
\sum_{d=0}^F p_{_{(I,m)}} = 1 \; .
\end{equation}

The states $(N,m)$ and $(E,m)$ are reached by transitioning from state $(I,m-1)$ with probabilities $\smash{q_{\gamma,m}}$ and $\smash{p_{\gamma,m}}$, respectively. Thus, we have $\smash{p_{_{(N,m)}} = q_{\gamma,m} p_{_{(I,m-1)}}}$ and $\smash{p_{_{(E,m)}} = p_{\gamma,m} p_{_{(I,m-1)}}}$, respectively, for $m = 1,\dots,F-1$. The final states $(N,F)$ and $(E,F)$ can be reached both from $(N,F-1)$, or $(E,F-1)$, and from $(N,F)$, or $(E,F)$, respectively. This gives us $\smash{p_{_{(N,F)}} = q_{\gamma,F} p_{_{(I,F-1)}} + q_{\gamma,f} p_{_{(I,F)}}}$ and $\smash{p_{_{(E,F)}} = p_{\gamma,F} p_{_{(I,F-1)}} + p_{\gamma,f} p_{_{(I,F)}}}$, respectively. The states $(T,m)$, are reached only from the event states $(E,m)$, and so we have $\smash{p_{_{(T,m)}} = p_\alpha p_{_{(E,m)}}}$. Note that a node in any of the $(T,m)$ states gets to transmit. The transmission probability of a node, denoted $\smash{p_{\mathrm{tx}} = \sum_{m=1}^F p_{_{(T,m)}}}$. A busy channel results when more than one such node accesses the channel at the same time. For a network with $M$ nodes, the $j^{th}$ node's probability of a busy channel is
\begin{equation}
\label{Eq:p}
p^{_{(j)}} = 1-\prod_{i \neq j,i=1}^{M} (1-p^{_{(i)}}_{\mathrm{tx}}) \; ,
\end{equation}
where $p_{\mathrm{tx}}^{_{(i)}}$ is the transmission probability of any of the other $M-1$ nodes. Note that we use the independence aspect of Assumption~\ref{Assumption:Bianchi} here, which simplifies the analysis.

For a network with $M$ nodes, we have $2M$ equations (\ref{Eq:sumb0}) and (\ref{Eq:p}) in $2M$ variables, $\smash{p^{_{(j)}}_{_{(I,0)}}}$ and $\smash{p^{_{(j)}}}$ for $j \in \{1,\dots,M\}$. These can be solved to find the corresponding steady state solution for each node in the network. Finally, a node that is successful in transmission, transitions to the state $(I,0)$. Thus, the probability of a successful transmission is given by $\smash{p_{_{(I,0)}}}$ in (\ref{Eq:PrDelta1}).
\end{IEEEproof}

The reliability is a joint measure of transmission obtained from the event-triggering policy and CRM. Other performance measures can also be found from the above Markov chain-based analysis. The steady state conditional probability of a successful transmission given that an event has occurred is given by $\lim_{k \rightarrow \infty} \Pr(\delta^{_{(j)}}_k = 1 | \gamma^{_{(j)}}_k=1) = p_{\alpha} q^{_{(j)}}$, which does not depend on the memory of the scheduler in steady state. This quantity measures the contribution of the CRM and other network traffic towards congestion, or the lack of it. Similarly, we can evaluate the delay distribution for a node in this network, as we show below.

\begin{corollary} \label{Corollary:DelayDist}
The delay distribution for a system described by (\ref{Eq:StateSpace})--(\ref{Eq:Controller}), with Assumption~\ref{Assumption:Bianchi} and $\zeta \in \mathbb{Z}$, is given by
\begin{equation} \label{Eq:DelayDist}
\Pr_d^{_{(j)}}(\zeta) = \begin{cases}
p^{_{(j)}}_{_{(I,d_k)}} p^{_{(j)}}_{\gamma,d_k} p_\alpha q^{_{(j)}} & d^{_{(j)}}_k < F \; , \\
\hat{p}^{_{(j)}}_{_{(I,d_k)}} p^{_{(j)}}_{\gamma,f} p_\alpha q^{_{(j)}} & d^{_{(j)}}_k \ge F \; , \end{cases}
\end{equation}
where $\hat{p}^{_{(j)}}_{_{(I,d_k)}} = (1 - p^{_{(j)}}_{\gamma,F} p_\alpha q^{_{(j)}}) p_{_{(I,F-1)}} + (1 - p^{_{(j)}}_{\gamma,f} p_\alpha q^{_{(j)}})^{(d_{k} - F)} p_{_{(I,F)}}$.
\end{corollary}
\begin{IEEEproof}
The probability of a delay $d < F$ is given by the probability of a successful transmission from the state $(T,d)$ to the state $(I,0)$, in Fig.~\ref{Fig:MCetCRM}. We use the same principle while computing the probability of a delay $d \ge F$. A delay of $d^{_{(j)}}_k = F$ is incurred when a successful transmission from $(T,F)$ is preceded by a transition from state $(I,F-1)$ to $(I,F)$. A delay of $d^{_{(j)}}_k > F$ is incurred when a successful transmission from $(T,F)$ is preceded by $d^{_{(j)}}_k - F$ transitions from state $(I,F)$ to itself and the aforemention transition from state $(I,F-1)$ to $(I,F)$. Using the expressions in (\ref{Eq:pIdleState}), we obtain (\ref{Eq:DelayDist}).
\end{IEEEproof}

Thus, the Markov model in Fig.~\ref{Fig:MCetCRM} helps us characterize the performance of the event-triggering policy and the CRM, for the entire network.

\subsection{An Event-triggering Policy as a Set of Probabilities}

In the Markov model presented in Fig.~\ref{Fig:MCetCRM}, the probability of an event $p_{\gamma,m}$ varies with $m$. This is because the input arguments to the policy, defined in (\ref{Eq:ETesterrip}), and the event thresholds, $\Delta$ in (\ref{Eq:InnoSched}), vary with $m$. Now, given that the event-triggering policy uses the estimation error as input, the set of thresholds $\{\Delta(0),\dots,\Delta(F)\}$, for $0 \le m \le F$, represent the chosen policy. This set of thresholds can be translated into the corresponding set of probabilities $\{p_{\gamma,1},\dots,p_{\gamma,F},p_{\gamma,f}\}$ using (\ref{Eq:TransProb_pgm}). Thus, the set of probabilities are an alternative representation of the chosen policy. In fact, the set of probabilities can represent any given event-triggering policy. Furthermore, this set of probabilities determines the performance of the event-based network, as we saw in Theorem~\ref{Theorem:SteadyStateAnalysis} and Corollary~\ref{Corollary:DelayDist}. Thus, we consider this set of probabilities as the specification of our event-triggering policy.

To implement a given event-triggering policy, a set of thresholds corresponding to the specified set of probabilities must be found. This is not a trivial task, as the prediction errors are not Gaussian. Furthermore, its probability densities are determined by the conditional probability of a busy channel $p$. However, it is worth noting that finding the set of probabilities corresponding to a set of thresholds is equally hard, as the underlying density functions need to be evaluated either way. It is easier to accomplish a translation from one representation to the other numerically, as we show in Section~\ref{S:Sims}.

\subsection{Extensions to More General Network Settings} \label{S:Extn}

We now extend the Markov model presented in Fig.~\ref{Fig:MCetCRM} to include more general network settings, such as retransmissions in the CRM and asynchrony. 

\subsubsection{CSMA with retransmissions}

\begin{figure*}[!t]
\begin{center}
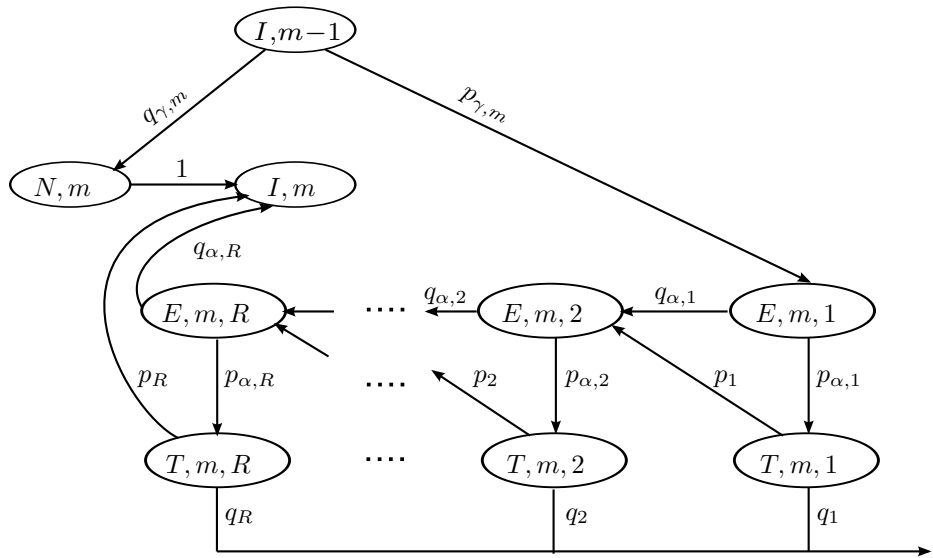
\caption{Embedding a CRM with $R$ distinct re-transmission stages in the Markov chain model} \label{Fig:MCretx}
\end{center}
\end{figure*}

A realistic CRM is likely to use retransmissions to spread congested network traffic over the sampling interval, as described in Fig.~\ref{Fig:TimeScales}. The model corresponding to such a CRM requires a Markov chain of its own, as shown in Fig.~\ref{Fig:MCretx}. Here, the event and CRM states, $(E,m)$ and $(T,m)$ for each $m$, are replaced by multiple states, $(E,m,r)$ and $(T,m,r)$, for $r=1,\dots,R$ retransmission attempts. Each successive retransmission attempt sees the same or lesser traffic from all the nodes in the network, due to a strictly non-negative probability of successful transmission in the previous attempt. Thus, the resulting Markov chain must have a unique conditional probability of a busy channel, $p_r$, for each retransmission attempt $1 \le r \le R$, analogous to $p$ in Assumption~\ref{Assumption:Bianchi}.
\begin{assumption} \label{Assumption:BianchiMulti}
For the systems described in (\ref{Eq:StateSpace})--(\ref{Eq:Controller}), the conditional probability of a busy channel for a node that attempts to transmit is given by an independent probability $p_r$ for each retransmission stage and each system. Thus,
\begin{equation} \label{Eq:BianchiMulti}
\lim_{k \to \infty} \Pr(\delta^{_{(j)}}_k=0|\gamma^{_{(j)}}_k=1,\alpha^{_{(j)}}_{k,r}=1) = p^{_{(j)}}_r \; ,
\end{equation}
for all $j \in \{1,\dots,M\}$ and all $r \in \{1,\dots,R\}$.
\end{assumption}

Generating a complete Markov chain for $m=0,\dots,F$, using the states shown in Fig.~\ref{Fig:MCretx}, we can re-evaluate all the probabilities in the proof of Theorem~\ref{Theorem:SteadyStateAnalysis}. Note that only some of the terms change. The probability of an unsuccessful transmission in the $m^{th}$ stage is now given by $\prod_{r=1}^{R} (p_{\alpha,r} p_r + q_{\alpha,r}) p_{\gamma,m}$, as follows from Fig.~\ref{Fig:MCretx}. This gives us
\begin{align*}
p_{_{(I,m)}} &= \prod_{r=1}^R (p_{\alpha,r} p_r + q_{\alpha,r}) p_{\gamma,m} p_{_{(I,m-1)}} \; , \quad m = 1,\dots,F-1 \; , \notag \\
p_{_{(I,F)}} &= \frac{\prod_{r=1}^R (p_{\alpha,r} p_r + q_{\alpha,r}) p_{\gamma,F}}{1-\prod_{r=1}^R (p_{\alpha,r} p_r + q_{\alpha,r}) p_{\gamma,f}} p_{_{(I,F-1)}} \; .
\end{align*}
The probability of the states $(T,m,r)$ is given by $p_{_{(T,m,r)}} = (\prod_{q=1}^{r-1} (p_{\alpha,q} p_q + q_{\alpha,q})) p_{\alpha,r} p_{\gamma,m} p_{_{(I,m-1)}}$, and the corresponding probability of transmission from any of the $(T,m,r)$ states, for different values of $r$, is given by $p_{\mathrm{tx},r} = \sum_{m=1}^F p_{_{(T,m,r)}}$. Now, the conditional probability of a busy channel in the $r^{\textrm{th}}$ retransmission stage can be derived as
\begin{equation} \label{Eq:pRetx}
p^{_{(j)}}_r = 1-\prod_{i \neq j,i=1}^{M} (1-p_{\mathrm{tx},r}^{_{(i)}}) \; , \text{for} \; r \in \{1,\dots,R\} \; .
\end{equation}
Using the above equations in place of (\ref{Eq:pIdleState}) and (\ref{Eq:p}), we obtain similar expressions for the reliability and delay distribution as in Theorem~\ref{Theorem:SteadyStateAnalysis} and Corollary~\ref{Corollary:DelayDist}, respectively.

We present simulations to validate Assumption~\ref{Assumption:BianchiMulti} in Section~\ref{S:Sims}.

\subsubsection{Asynchronous networks}

\begin{figure*}[!t]
\begin{center}
\def\svgwidth{10cm}
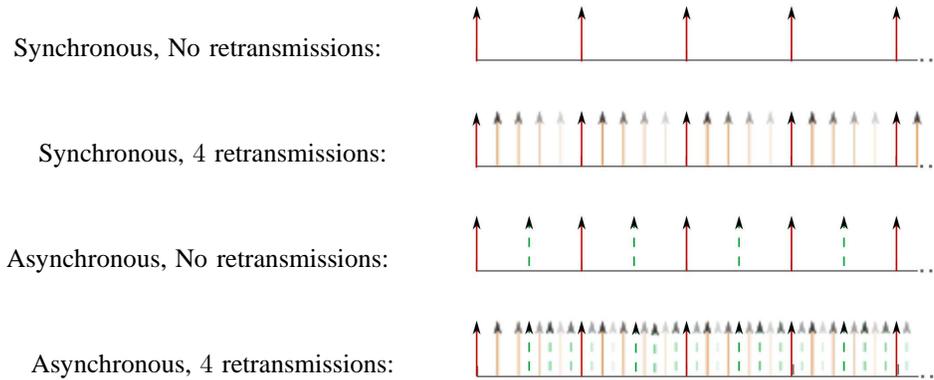
\caption{A comparison between synchronous and asynchronous traffic, with and without retransmissions in the CRM. The steady state analysis differs for each traffic pattern, as nodes in the $(T,m)$ or $(T,m,r)$ states see different traffic patterns under each configuration. } \label{Fig:AsyncRetx}
\end{center}
\end{figure*}

Consider an asynchronous network, with the CRM operating in a beacon-enabled mode. In this mode, the CRM slots remain synchronized across the network, but different systems can choose to initiate sampling at randomly selected CRM slots. Consecutive samples are spaced by the sampling period $T$ CRM slots, for all the systems in the network. An illustration of the behaviour, with and without retransmissions in the CRM, for synchronous and asynchronous networks, is provided in Fig.~\ref{Fig:AsyncRetx}. For an asynchronous network with no retransmissions in the CRM, the number of interfering transmissions in the $(T,m)$ states is given by $M^{(j)} < M$, where $M^{(j)}$ is the number of nodes whose sampling instants lie in the same MAC slot of the $j^{th}$ node. Thus, the performance of the network depends on the initial sampling slots chosen by the nodes. The more spread apart they are, the better the performance. For an asynchronous network with retransmissions in the CRM, the steady state seen by each retransmission state $(T,m,r)$ can only be determined by knowing which of the retransmission states of other nodes interferes with transmissions from the $r^{\textrm{th}}$ stage. Thus, to predict the performance of such a network, one must know the initial sampling slots chosen by all the nodes in the network.

If we assume that the initial sampling slots are chosen uniformly across a frame, we can predict the average performance of an asynchronous network. We can compute the conditional probability of a busy channel by averaging across all possible combinations of interactions in each retransmission state. This averaging makes a node in the state $(T,m,r)$ see a busy channel due to other nodes in any of their $(T,m,r)$ states, for all $m$ and $r$. Thus, the probability of a busy channel is uniform across all retransmission states, i.e., $p^{_{(j)}}_r = p^{_{(j)}}$ for $r=1,\dots,R$. The modified version of Assumption~\ref{Assumption:Bianchi} is stated below.
\begin{assumption} \label{Assumption:BianchiAsync}
For the systems described in (\ref{Eq:StateSpace})--(\ref{Eq:Controller}) in an asynchronous network, the conditional probability of a busy channel for a node that attempts to transmit, is given by an independent probability $p$ for all retransmission stages, for each system. Thus,
\begin{equation} \label{Eq:BianchiAsync}
\lim_{k \to \infty} \Pr(\delta^{_{(j)}}_k=0|\gamma^{_{(j)}}_k=1,\alpha^{_{(j)}}_{k,r}=1) = p^{_{(j)}} \; ,
\end{equation}
for all $j \in \{1,\dots,M\}$ and all $r \in \{1,\dots,R\}$.
\end{assumption}

To evaluate the probability in \eqref{Eq:BianchiAsync}, we average across all the competing transmissions during the slots corresponding to state $(T,m,r)$ of the $j^{\textrm{th}}$ node, for some $m$ and $r$. There are $R^{M-1}$ different combinations of interactions between the $R$ retransmission stages of the other $M-1$ nodes in the network, due to different initial sampling slots. If each of these interactions are equally likely, the resulting expression is quite simple. The conditional probability can be found to be
\begin{equation} \label{Eq:pAsync}
p^{_{(j)}} = 1-\prod_{i \neq j,i=1}^{M} (1-q_{\mathrm{tx}}^{_{(i)}}) \; ,
\end{equation}
where $q_{\mathrm{tx}}^{_{(i)}} = (1/R) \cdot \sum_{r=1}^R q_{\mathrm{tx},r}^{_{(i)}}$ denotes the average transmission probability across all retransmission states. This equation can be used in place of (\ref{Eq:p}) to find expressions for the reliability and delay distribution as before. We perform simulations in Section~\ref{S:Sims} to validate Assumption~\ref{Assumption:BianchiAsync} and the resulting analysis. However, note that to obtain this result, we simulate across $R^{M-1}$ different combinations of interactions, due to $R^{M-1}$ different combinations of initial sampling slots. The result obtained for a single selection of sampling slots can be quite different from the averaged values.

\section{Examples and Simulations} \label{S:Sims}

We now return to Example~\ref{Ex:Setup}, and apply our analysis to this experimental setup. We present a number of variations of this example to validate each of the assumptions we have used for analyzing different network configurations. In each case, we present the reliability obtained through Monte-Carlo simulations, and compare it to the analytical value obtained using the analysis presented above. The differences are negligible in each case, thus validating our assumptions and verifying our analysis. We also evaluate the Linear Quadratic Gaussian (LQG) control cost, defined as
\begin{equation*}
\lim_{N \to \infty} \frac{1}{N} \sum_{n=0}^{N-1} \E \left[ x_n^T Q_1 x_n + u_n^T Q_2 u_n \right] \; ,
\end{equation*}
where $Q_1$ and $Q_2$ are the state and control weighting matrices, respectively. We use the LQG cost as a control-theoretic performance metric for a given event-triggering policy.

\begin{example}[Event-Triggering Policy as a Set of Probabilities] \label{Ex:Setup2}
We use the same setup described in Example~\ref{Ex:Setup}, comprising of a homogenous network of $M=10$ scalar systems, with $R=5$ retransmissions. The event-triggering policies are described by \eqref{Eq:InnoSched}, with constant thresholds. For a chosen set of event probabilities, we discuss the implementation of the event-triggering policy. We also compare the results of Monte-Carlo simulations with results obtained from our analysis.

\begin{table}[!t]
\begin{center}
\caption{A comparison of analytical and simulated values of $p$}\label{Tb:SimResults}
\begin{tabular}{c | c c}
\hline\hline Parameter & Simulation & Analysis \\ \hline
$p_\delta$ & $0.1840$ & $0.1872$ \\
$p_1$ & $0.5937$ & $0.5944$ \\
$p_2$ & $0.5655$ & $0.5620$ \\
$p_3$ & $0.5367$ & $0.5277$ \\
$p_4$ & $0.5076$ & $0.4917$ \\
$p_5$ & $0.4778$ & $0.4542$ \\
\hline
\end{tabular}
\end{center}
\vspace{-2mm}
\end{table}

The event probabilities are given to be $p_{\gamma,m} = \begin{bmatrix} 0.3171 & 0.5138 \end{bmatrix}$ for $m = 1,\dots,M$. Computing thresholds from a set of event probabilities is not easy, as the estimation error distributions are not Gaussian. In fact, a closed-form expression cannot be found for the distribution, though the evolution of the distribution can be described iteratively. Thus, we empirically select thresholds which result in the desired probabilities. In fact, $\Delta = 1$ achieves the given probabilities.

The values of reliability and the conditional probabilities of a busy channel obtained through simulations and analysis are presented in Table~\ref{Tb:SimResults}. The simulated values agree closely with the analytical values computed using Theorem~\ref{Theorem:SteadyStateAnalysis}. Thus, Assumption~\ref{Assumption:BianchiMulti} is a reasonable approximation and motivates the Markov modelling.
\end{example}

\begin{example}[Simple Network with No Retransmissions] \label{Ex:NoRetx}
We now consider a setup consisting of $M=2$ nodes. There are no retransmissions in this network, i.e., $R=1$, and the CRM probability is $p_{\alpha}=0.5$. The plant model and event-triggering policy are identical to the ones in Example~\ref{Ex:Setup}. The scheduler threshold is varied from $\Delta = 0$ to $\Delta = 8$ in this experiment. Each value of $\Delta$ results in a set of event probabilities and a corresponding network performance. The reliability $p_\delta$ obtained through analysis and simulations is plotted against the threshold in Fig.~\ref{Fig:pDeltaVSeps_NoRetx}. As the threshold increases, the reliability decreases. Thus, at larger thresholds, too few events are being generated to completely utilize the network resources. Note the close correspondence between the simulated and analytical values, validating Assumption~\ref{Assumption:Bianchi}. The corresponding control costs obtained through simulations are also plotted in the graph below, indicating an expected increase in cost with decreasing reliability.

\begin{figure*}[tb]
\begin{center}
\includegraphics*[scale=0.35, viewport = 80 55 1400 800]{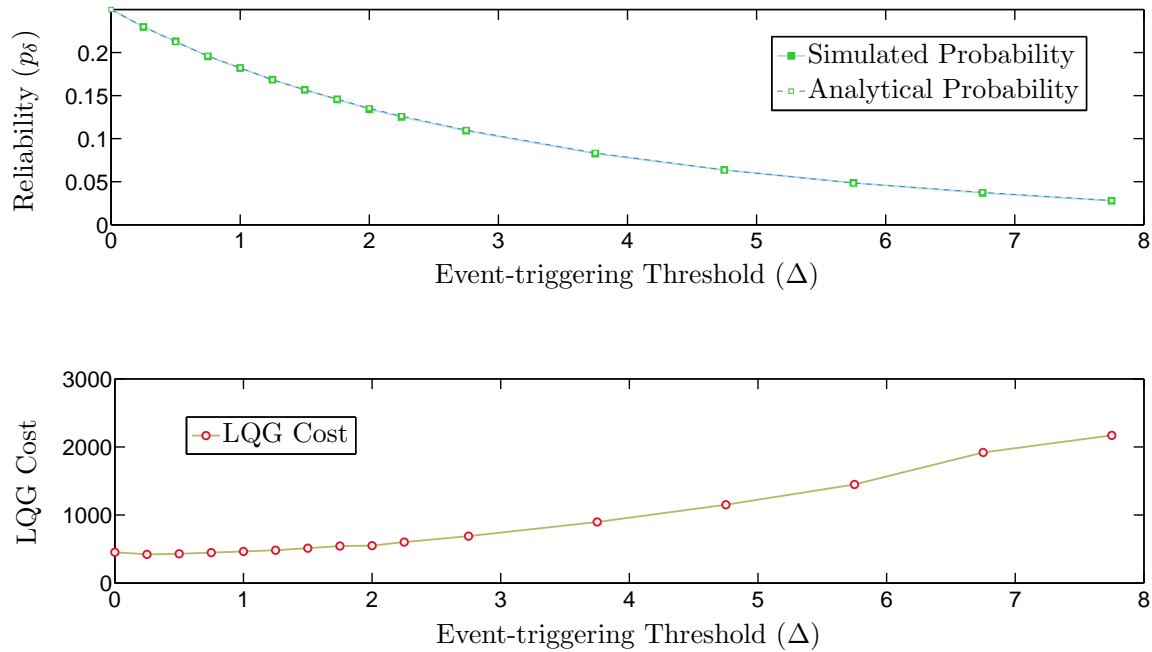}
\caption{A comparison of the analytical and simulated values of the reliability versus the event-triggering threshold, in a simple network with no retransmissions in the CRM. The close correspondence of these values validates Assumption~\ref{Assumption:Bianchi} and the results of Theorem~\ref{Theorem:SteadyStateAnalysis}.} \label{Fig:pDeltaVSeps_NoRetx}
\end{center}
\end{figure*}
\end{example}

\begin{example}[Retransmissions in the CRM] \label{Ex:Retx}
We return to the problem setup in Example~\ref{Ex:Setup}, with $M=10$ nodes and $R=5$ retransmissions. A comparison of analytical and simulated values of the reliability versus the threshold for this synchronized network is shown in Fig.~\ref{Fig:pDeltaVSepsilon}. The performance obtained from the network is, in accordance with expectations, poor due to synchronization and congestion. Low thresholds cause many packets to flood the network, and result in a low probability of a successful transmission due to congestion. High thresholds reduce the utilization of the network, and the probability of a successful transmission decreases again. Note that there is a threshold that optimizes use of the network resources. A system-level performance analysis is required to characterize this threshold.

\begin{figure*}[tb]
\begin{center}
\includegraphics*[scale=0.35, viewport = 80 55 1400 800]{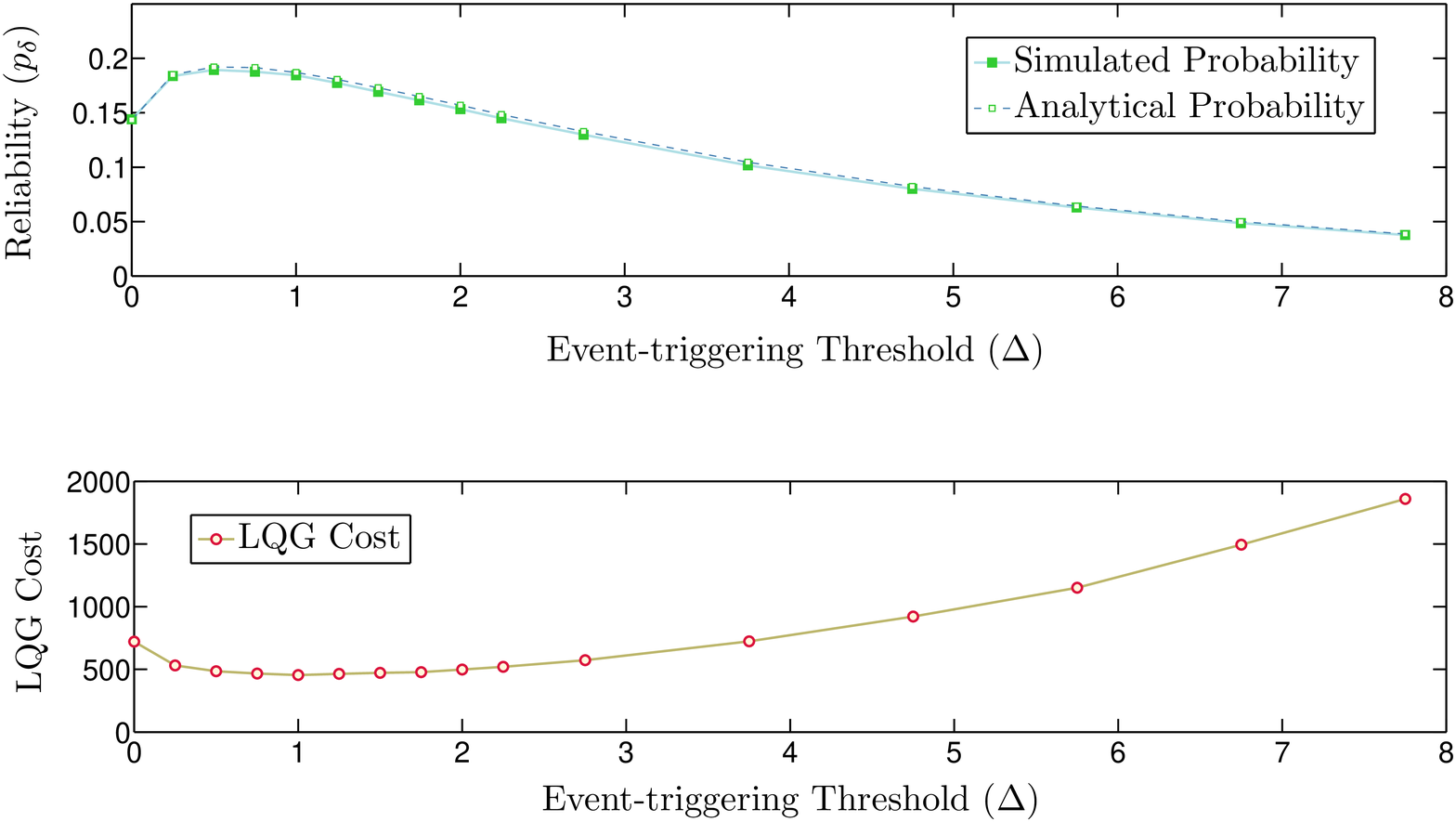}
\caption{A comparison of the analytical and simulated values of the reliability versus the scheduler threshold, with retransmissions in the CRM. This example validates Assumption~\ref{Assumption:BianchiMulti}. Note that low thresholds result in a low $\Pr(\delta_k = 1)$ due to congestion. High thresholds also result in a low $\Pr(\delta_k = 1)$, but due to under-utilization of the network. } \label{Fig:pDeltaVSepsilon}
\end{center}
\end{figure*}
\end{example}

\begin{example}[Unsaturated Traffic] \label{Ex:Unsat}
In this example, we look at sparse traffic and show that Bianchi's assumption holds well even in this scenario. We have now validated Bianchi's assumption in two different scenarios, with and without retransmissions in the CRM. However, Bianchi's assumption is theoretically motivated by a mean field analysis. Thus, it is important to ascertain that this assumption holds just as well when there is not much traffic in the network. So, consider a network with $M=2$ nodes and $R=5$ retransmissions in the CRM. The sampling period corresponds to $T=5$ CRM slots. Thus, each of the nodes has sufficient slots to transmit successfully. The plant model and event-triggering policy are the same as in Example~\ref{Ex:Setup}.

A comparison of the reliability obtained for different thresholds is shown in Fig.~\ref{Fig:pDeltaVSeps_unsat}. Note that the maximum reliability obtained in this network is for the lowest value of the threshold, i.e., $\Delta=0$. In other words, all samples are chosen as events, and even so, the network is largely successful in delivering them to the respective controllers. Also note that the reliability falls sharply as the threshold increases, indicating that too few events are generated to fully utilize the available network resources.

\begin{figure*}[tb]
\begin{center}
\includegraphics*[scale=0.35, viewport = 80 55 1400 800]{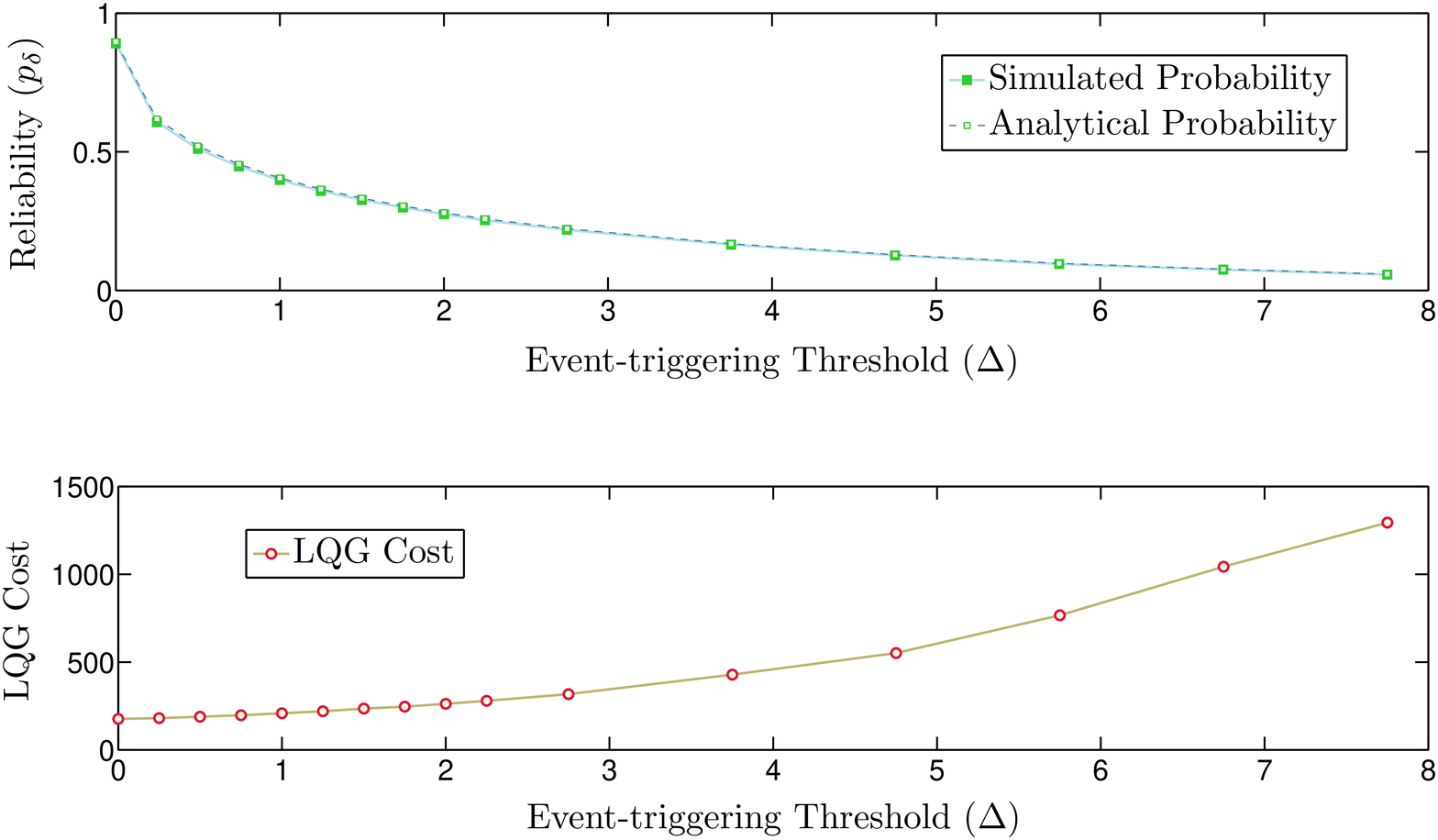}
\caption{A comparison of the analytical and simulated values of the reliability versus the scheduler threshold, with unsaturated traffic. There are just $2$ nodes in the network and the CRM permits $5$ retransmissions. Even so, Bianchi's assumption seems to hold, indicating that this is a good approximation of unsaturated and saturated network conditions.} \label{Fig:pDeltaVSeps_unsat}
\end{center}
\end{figure*}
\end{example}

\begin{example}[Asynchronous Traffic] \label{Ex:Async}
We now look at an asynchronous network, with $M=5$ nodes, a sampling period of $T=3$ slots and $R=2$ retransmissions in the CRM. The plant model and event-triggering policy are identical to the setup in Example~\ref{Ex:Setup}. The persistence probabilities of the CRM are chosen to be $p_{\alpha,1} = p_{\alpha,2} = 0.4$. A comparison of the reliabilities obtained for various thresholds is shown in Fig.~\ref{Fig:pDeltaVSeps_Async}. The analytical and simulated values bear close correspondence, thus validating Assumption~\ref{Assumption:BianchiAsync}. Note that the values obtained in this experiment are averaged across all possible selections of initial sampling slots by all five nodes in the network. The results may be quite different for a given selection of initial sampling slots. In other words, the highest reliability obtainable from this system may far exceed the average reliability shown in Fig.~\ref{Fig:pDeltaVSeps_Async}.
\begin{figure*}[tb]
\begin{center}
\includegraphics*[scale=0.35, viewport = 80 55 1400 800]{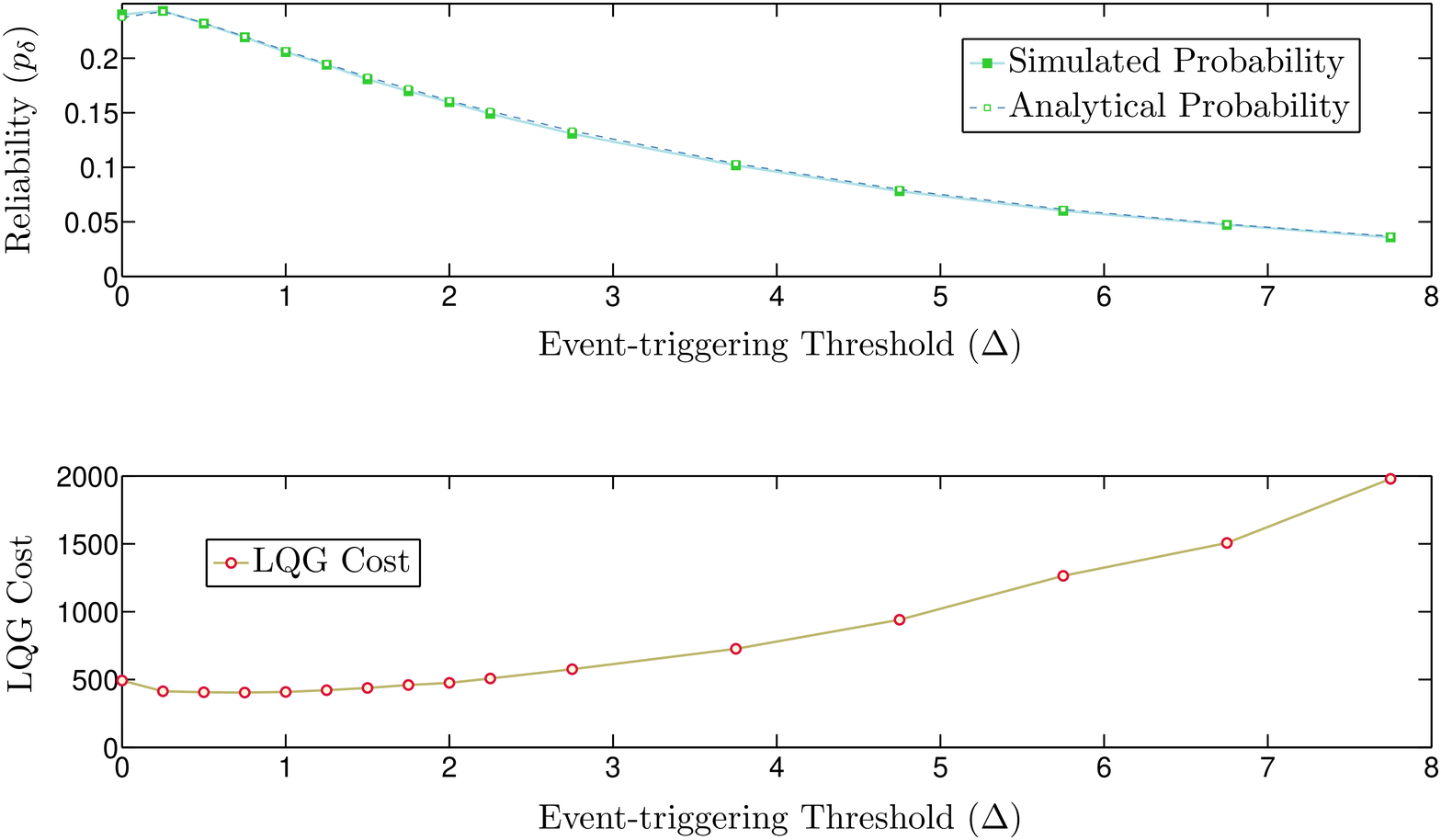}
\caption{A comparison of the analytical and simulated values of the reliability versus the event-triggering threshold, for an asynchronous network. The network consists of $5$ nodes, with sampling periods of $3$ slots and $2$ retransmissions in the CRM. The reliability obtained through analysis and simulations is averaged across all possible selections of initial sampling slots by the $5$ nodes in the network. This plot validates Assumption~\ref{Assumption:BianchiAsync}.} \label{Fig:pDeltaVSeps_Async}
\end{center}
\end{figure*}
\end{example}


\subsection{Discussion}

Let us now compare the results we have obtained in the above examples, to comment on the underlying network configurations. Examples~\ref{Ex:NoRetx}~and~\ref{Ex:Unsat} deal with networks consisting of two closed-loop systems each, but permitting one and five re-transmissions, respectively. The higher number of retransmissions results in a significantly higher reliability, and correspondingly lower LQG cost. This can be seen by comparing Figures~\ref{Fig:pDeltaVSeps_NoRetx}~and~\ref{Fig:pDeltaVSeps_unsat}. Example~\ref{Ex:Retx} deals with a network consisting of ten synchronized closed-loop systems and five retransmissions. The ratio of transmissions slots to number of systems is equal to $0.5$, which is the same as for Example~\ref{Ex:NoRetx}. A comparison of Figures~\ref{Fig:pDeltaVSepsilon}~and~\ref{Fig:pDeltaVSeps_NoRetx} indicates a slightly reduced reliability in the multiple retransmission case, especially for small values of the event-triggering threshold $\Delta$. This can be attributed to the increase in congestion at every sampling instant in a synchronized network with more systems. The reliability curve for the asynchronous network in Fig.~\ref{Fig:pDeltaVSeps_Async} improves the performance for small values of $\Delta$.

We now comment on Bianchi's assumption, which has been shown to hold under different network configurations. The above results validate the use of Bianchi's assumption for modelling the interactions of event-triggering policies and CRMs. Previously, Bianchi's assumption has been shown to hold in setups where the probability of accessing the network in different stages results from independent random processes, such as random backoffs in CSMA/CA. The probability of accessing the network in our model is not independent in each stage, as the estimation error for event-triggering policies is correlated to its past, as shown in Lemma~\ref{Lemma:LostProp}. Thus, what we have here is a new configuration for the applicability of Bianchi's assumption. A theoretical motivation of this assumption is beyond the scope of this paper. An explanation of Bianchi's assumption in the context of CSMA/CA is presented in \cite{Bordenave2010}.

\section{Conclusions} \label{S:Concl}

We have presented a method to analyze the performance of a network of event-based systems that use a CRM to access the shared network. We have shown that a Markov model can be constructed to represent the event-triggering policy and CRM, once we use Bianchi's assumption. Based on this model, we have analyzed the steady state performance of the resulting network. This analysis assumed conditional independence from other traffic when a node attempts to transmit. We validated this assumption through simulations, and provided extensions to more complex network configurations. For future work, we wish to use the insights obtained through this work to design adaptive event-triggering policies.

\bibliographystyle{ieeetr}
\bibliography{FullList}


\end{document}